\documentclass[aps,prb,twocolumn]{revtex4}
\usepackage{amsfonts}
\usepackage{amssymb}
\usepackage{amsmath}
\usepackage{graphicx}
\usepackage{epsfig}
\usepackage{color}

\begin{document}

\title{Non-Hermitian Floquet topological phases: Exceptional points,
coalescent edge modes, and the skin effect}
\author{Xizheng Zhang}
\affiliation{Department of Physics, National University of Singapore, Singapore 117551,
Republic of Singapore}
\author{Jiangbin Gong}
\email{phygj@nus.edu.sg}
\affiliation{Department of Physics, National University of Singapore, Singapore 117551,
Republic of Singapore}

\begin{abstract}
Periodically driven non-Hermitian systems can exhibit rich topological band
structure and non-Hermitian skin effect, without analogs in their static or
Hermitian counterparts. In this work we investigate the exceptional
band-touching points in the Floquet quasi-energy bands, the topological
characterization of such exceptions points and the Floquet non-Hermitian
skin effect (FNHSE). Specifically, we exploit the simplicity of periodically
quenched two-band systems in one dimension or two dimensions to analytically
obtain the Floquet effective Hamiltonian as well as locations of the many
exceptional points possessed by the Floquet bulk bands. Two different types
of topological winding numbers are used to characterize the topological
features. Bulk-edge correspondence (BBC) is naturally found to break down
due to FNHSE, which can be drastically different among different bulk
states. Remarkably, given the simple nature of our model systems, recovering
the BBC is doable in practice only for certain parameter regime where a
low-order truncation of the characteristic polynomial (which determines the
Floquet band structure) becomes feasible. Furthermore, irrespective of which
parameter regime we work with, we find a number of intriguing aspects of
Floquet topological zero modes and $\pi$ modes. For example, under the open
boundary condition zero edge modes and $\pi$ edge modes can individually
coalesce and localize at two different boundaries. These anomalous edge
states can also switch their accumulation boundaries when certain system
parameter is tuned. These results indicate that non-Hermitian Floquet
topological phases, though more challenging to understand than their
Hermitian counterparts, can be extremely rich in the presence of FNHSE.
\end{abstract}

\maketitle


\section{Introduction}

Periodic driving, when applied to spatially periodic systems, can yield rich
topological phases of matter, now often termed as Floquet topological matter
\cite{QC,Butterfly, Oka,Rudner1,PRE2013,Jiang,Bermudez,Gong,Ho,
An1,Rudner2,Rechtsman,Steinberg, Kundu,Miyake,Aidelsburger,Jotzu,Rudner3,An2,Anisimovas1,Hu,Anisimovas2,Gao,Eckardt,Mukherjee1,Maczewsky,Anisimovas3,Mahmood,WangB,Fertig,Seradjeh,Mukherjee2,Longwen,ChinghuaPRL2018,Asb1,Asb2,Bomantara,Chen,Zhou}%
. Edge states of Floquet topological matter are pinned at special
quasienergy values and the associated anomalous edge states \cite%
{Asb1,Asb2,Ho,Bomantara,Chen,Zhou} may not obey the usual bulk boundary
correspondence (BBC) \cite{Rudner2}. The classification of Floquet
topological phases has been established \cite{Roy,WangZ1} and corresponding
topological states have been observed in cold atom, photonic, phonoic and
acoustic systems \cite%
{Rechtsman,Jotzu,Steinberg,Mahmood,Miyake,Mukherjee1,Maczewsky,Mukherjee2,Kitagawa,Tarnowski,Fleury,Peng}%
. Effects of interaction and disorder on Floquet topological phases have
also been explored theoretically \cite%
{Rudner2,Titum,Khemani,Sreejith1,Sreejith2}.  In addition to the generation of novel topological
phases, periodic time modulation as a control protocol is also of great
interest to the realization of nonreciprocal propagation \cite%
{Rechtsman,Fang} and the design of nonreciprocal devices for photonic and
acoustic applications.

In a realistic experimental setup, a quantum system is likely to interact
with its environment. To account for this, certain non-Hermitian terms can
be introduced in our theoretical modelling. Over recent
years, non-Hermitian systems have gained a great deal of attention,
especially due to rapid progresses in experimental implementations of
non-Hermiticity. Such experiments include examples from photonics \cite%
{Poli,Zeuner,XuePeng1,XuePeng2,Weimann,Parto,Joannopoulos,Bandres,XuePeng3,Ozawa}%
, acoustics \cite{ChenH}, vacancy centers in solids \cite{DuJF}, and cold
atoms \cite{Luo}, \ where non-Hermiticity were introduced through
judiciously incorporating gain and loss \cite%
{Makris,Lin,Regensburger,PengBo,Hodaei}. Not surprisingly then, these
available experimental setups therefore make concrete realizations of
non-Hermitian lattice model possible, such as a non-Hermitian version of the
topological Su-Schrieffer-Heeger model \cite{Weimann,Zeuner,Schomerus,Ghatak}.
Non-Hermitian topological phases often exhibit significantly different
physics from their parent Hermitian counterparts \cite%
{SHT,Kunst,Kawabata1,Kawabata2,Jin,LiuT,ChingHuaPRL,GhatakJPC,LinhuPRB,Ghatak}. A
well-known example is the emergence of exceptional point (EP), where a
spectral degeneracy is accompanied by a coalescence of the corresponding
eigenstates \cite{Heiss}. Another remarkable feature is the stark difference
between the band spectra under periodic boundary condition and that under
open boundary condition (hereafter PBC and OBC, respectively). In
particular, the bulk eigenstates under OBC are generally localized near
boundaries, a phenomenon termed the non-Hermitian skin effect (NHSE) \cite%
{Yao1,Yao2,Chinghua,Chinghuaaxxiv}. It is now widely accepted that the
presence of NHSE often signifies a breakdown of the bulk-boundary
correspondence (BBC), {a concept already experimentally verified \cite{Helbig,Ghatak,Xuepeng2019,Hofmann}.}

Given the above two fruitful topics, namely, Floquet topological phases and
non-Hermitian topological phases, we are motivated to examine how
non-Hermiticity impacts on Floquet topological phases, especially the
interplay of NHSE and periodic driving. Specifically, we naturally ask the
following questions: (i) What is the main effect of non-Hermiticity on
Floquet-Bloch band structure? (ii) What will be the interesting features of
a Floquet system with NHSE? (iii) Do there exist EPs in the quasi-energy
spectrum of a periodically driven system? If yes, what is the topology of
the EPs? (iv) What is the destiny of BBC in non-Hermitian Floquet systems?
In this work, we attempt to answer these questions by focusing on two
concrete non-Hermitian two-band systems subject to periodic quenching \cite{An2,ZhoulPRB,Zhou}. One
main reason to choose such systems is that the Floquet effective Hamiltonian
can be easily obtained.

A number of findings are in order. (i) A periodically quenched system can
exhibit the skin effect (dubbed as Floquet non-Hermitian skin effect
(FNHSE)) even though the quenched Hamiltonian in each step does not have
NHSE. (ii) The introduction of a non-Hermitian term can not only split
spectral degenerate point (DP) of the parent Hermitian Floquet system into
two EPs but also induce many other EPs. Such EPs are unique insofar as they
each carry half-integer topological winding numbers. These results were
totally absent in a previously studied non-Hermitian Floquet system that
does not possess FNHSE \cite{ZhoulPRB}. (iii) The presence of FNHSE breaks
the BBC. However, within certain parameter regime, the existence of two
different types of Floquet edge modes can still be predicted exactly by
introducing the generalized Brillouin zone (GBZ) in two time-symmetric
frames. Moreover, FNHSE can either push the edge modes to one side of the
system or separate two different types of Floquet edge modes at two opposite
boundaries. It is hoped that these results can motivate further studies of
both fundamental aspects and potential applications of non-Hermitian Floquet
topological phases.

The rest of this paper is organized as follows: In Sec.~\ref{general}, we
outline a general treatment of two-step quenched {non-Hermitian Floquet
systems and introduce two types of winding numbers for topological
characterization of EPs. With these preparations, in Sec.~\ref{OBC} we study
a two-dimensional (2D) model, with an emphasis placed on topological
characterization of the EPs in the PBC spectrum. To better digest FNHSE, we
treat a one-dimensional (1D) model in Sec.~\ref{1D}. There we examine the
possibility of recovering the BBC with certain parameter regime. Sec.~\ref%
{conclusions} concludes this paper. Some nonessential details of our
calculation are placed in Appendix. \ }

\section{Time-periodic quenching and topological characterization under PBC}

\label{general}

\subsection{Floquet bands}

Periodic driving such as periodic quenching is capable of producing
topological phases with counter-intuitively large topological invariants
\cite{Longwen,ZhoulPRB}. One qualitative explanation is that periodic driving can
effectively induce long-range hopping \cite{An1,An2,Longwen}. With this
physical picture in mind, we anticipate that the interplay of
non-Hermiticity and periodic driving can be highly nontrivial. For
simplicity we focus on periodic quenching applied to physically realistic
non-Hermitian tight-binding lattice models with only nearest-neighbor (NN)
hoppings.

Consider then a periodically driving protocol with overall driving period $%
T=T_{1}+T_{2}$, such that $H=H_{1}$ for duration $T_{1}$ and $H=H_{2}$ for
duration $T_{2}$. We further confine our discussions to simple two-band
dimensionless (with $\hbar =1$) Hamiltonians that only contain Pauli matrix
but are enough to produce rich topological phases. That is, we assume
\begin{equation}
H_{i\text{ }\left( i=1,2\right) }=\sum_{\mathbf{k}}\mathbf{B}_{i}\left(
\mathbf{k}\right) \cdot \mathbf{\sigma }\left\vert \mathbf{k}\right\rangle
\left\langle \mathbf{k}\right\vert ,
\end{equation}%
where $\mathbf{k\in (}-\pi ,\pi ]$ is the quasimomentum vector, and $\mathbf{%
B}_{i}\left( \mathbf{k}\right) =\mathbf{h}_{i}\left( \mathbf{k}\right) +i%
\mathbf{g}_{i}\left( \mathbf{k}\right) $ with $\mathbf{h}_{i}\left( \mathbf{k%
}\right) $ and $\mathbf{g}_{i}\left( \mathbf{k}\right) $ being a real vector
function, with their components attached to different Pauli matrices. The
non-Hermiticity of $H_{i}$ stems from the non-Hermitian term $i\mathbf{g}%
_{i}\left( \mathbf{k}\right) $, which can represent either on-site complex
potential or non-reciprocal hopping along a lattice. Regarding possible
experimental setups, ultracold atomic gas in optical lattices, photonic
crystals, and coupled resonators provide versatile platforms to realize such
non-Hermitian systems, with tunability in the system parameters. For
example, the non-Hermiticity may be realized by asymmetric scattering
between a clockwise and a counterclockwise propagating mode within each
resonator, or by introducing atom loss (to effectively induce some
non-reciprocal hopping) in the cold-atom context. Given the realizations of $%
H_1$ and $H_2$, the periodic quenching between them further requires
periodic modulation of certain system parameters (e.g., the depth of optical
lattice potentials for a platform involving cold-atoms and optical lattice
potentials). Alternatively, quantum walk experiments \cite{Xuepeng2019} also
provide a natural platform for realizing periodically quenched quantum
systems.

Our general consideration starts from the following single-period Floquet
operator:
\begin{equation}
\widehat{U}_{T}=\sum_{\mathbf{k}}U\left( \mathbf{k}\right) \left\vert
\mathbf{k}\right\rangle \left\langle \mathbf{k}\right\vert ,
\end{equation}%
with
\begin{equation}
U\left( \mathbf{k}\right) =e^{-iT_{2}\mathbf{B}_{2}\left( \mathbf{k}\right)
\cdot \mathbf{\sigma }}e^{-iT_{1}\mathbf{B}_{1}\left( \mathbf{k}\right)
\cdot \mathbf{\sigma }},
\end{equation}
{which defines an effective Floquet Hamiltonian $H_{\mathrm{eff}}\left(
\mathbf{k}\right) =i\ln U\left( \mathbf{k}\right) /T$}. The corresponding
Floquet eigenstates of $U\left( \mathbf{k}\right) $ are obtained from
\begin{equation}
U\left( \mathbf{k}\right) \left\vert \varphi \right\rangle =e^{-i\varepsilon
\left( \mathbf{k}\right) T}\left\vert \varphi \right\rangle ,
\end{equation}%
where $\varepsilon \left( \mathbf{k}\right) $ are the quasienergies and $%
\left\vert \varphi \right\rangle $ are the corresponding Floquet
eigenstates. Due to the non-Hermiticity of $H_{i}$, the Floquet operator is
not unitary and therefore $\varepsilon \left( \mathbf{k}\right) $ is usually
not real. Nevetheless, the real parts of $\varepsilon \left( \mathbf{k}%
\right) $ can still be deemed as a phase defined up to a multiple of $2\pi
/T $, generally chosen to lie in the first quasienergy Brillouin zone $%
\mathbf{(}-\pi /T,\pi /T]$.

The closed $SU(2)$ algebra here provides an explicit form of $U\left(
\mathbf{k}\right) =\exp \left[ -i\varepsilon \left( \mathbf{k}\right)
\widehat{\mathbf{n}}\left( \mathbf{k}\right) \cdot \mathbf{\sigma }\right] $%
, where $\widehat{\mathbf{n}}=\mathbf{n}\left( \mathbf{k}\right) /\left\Vert
\mathbf{n}\left( \mathbf{k}\right) \right\Vert $ is in general a complex
unit vector with
\begin{eqnarray}
\mathbf{n\left( \mathbf{k}\right) } &\mathbf{=}&\sin \left( \left\Vert
\mathbf{B}_{2}\left( \mathbf{k}\right) \right\Vert T_{2}\right) \cos \left(
\left\Vert \mathbf{B}_{1}\left( \mathbf{k}\right) \right\Vert T_{1}\right)
\widehat{\mathbf{B}}_{2}\left( \mathbf{k}\right)  \notag \\
&&+\cos \left( \left\Vert \mathbf{B}_{2}\left( \mathbf{k}\right) \right\Vert
T_{2}\right) \sin \left( \left\Vert \mathbf{B}_{1}\left( \mathbf{k}\right)
\right\Vert T_{1}\right) \widehat{\mathbf{B}}_{1}\left( \mathbf{k}\right)
\notag \\
&&+\left[ \sin \left( \left\Vert \mathbf{B}_{1}\left( \mathbf{k}\right)
\right\Vert T_{1}\right) \sin \left( \left\Vert \mathbf{B}_{2}\left( \mathbf{%
k}\right) \right\Vert T_{2}\right) \right.  \notag \\
&&\widehat{\mathbf{B}}_{2}\left( \mathbf{k}\right) \times \widehat{\mathbf{B}%
}_{1}\left( \mathbf{k}\right) ].
\end{eqnarray}%
The expression of the quasienergies $\varepsilon \left( \mathbf{k}\right) $
can be given as
\begin{eqnarray}
\cos \left[ \varepsilon \left( \mathbf{k}\right) \right] &=&\cos \left(
\left\Vert \mathbf{B}_{1}\left( \mathbf{k}\right) \right\Vert T_{1}\right)
\cos \left( \left\Vert \mathbf{B}_{2}\left( \mathbf{k}\right) \right\Vert
T_{2}\right)  \notag \\
&&-\left[ \sin \left( \left\Vert \mathbf{B}_{1}\left( \mathbf{k}\right)
\right\Vert T_{1}\right) \sin \left( \left\Vert \mathbf{B}_{2}\left( \mathbf{%
k}\right) \right\Vert T_{2}\right) \right.  \notag \\
&&\times \widehat{\mathbf{B}}_{1}\left( \mathbf{k}\right) \cdot \widehat{%
\mathbf{B}}_{2}\left( \mathbf{k}\right) ]  \label{quasi_energy}
\end{eqnarray}%
with $\left\Vert \mathbf{B}_{i}\left( \mathbf{k}\right) \right\Vert =\sqrt{%
B_{i,x}^{2}\left( \mathbf{k}\right) +B_{i,y}^{2}\left( \mathbf{k}\right)
+B_{i,z}^{2}\left( \mathbf{k}\right) }$ and $\widehat{\mathbf{B}}_{i}\left(
\mathbf{k}\right) =\mathbf{B}_{i}\left( \mathbf{k}\right) /\left\Vert
\mathbf{B}_{i}\left( \mathbf{k}\right) \right\Vert $ are also the complex
unit vectors. From the expression of $\varepsilon \left( \mathbf{k}\right) $
given above, it can be seen that there may exist two quasienergy gaps at $%
\varepsilon \left( \mathbf{k}_{c}\right) =0$ and $\varepsilon \left( \mathbf{%
k}_{c}\right) =\pi $ in the complex band structure. The band touching points
are located at $\varepsilon \left( \mathbf{k}_{c}\right) =0$ and $%
\varepsilon \left( \mathbf{k}_{c}\right) =\pi $, equivalent to two criteria $%
\cos \left[ \varepsilon \left( \mathbf{k}_{c}\right) \right] =1$ or $\cos %
\left[ \varepsilon \left( \mathbf{k}_{c}\right) \right] =-1$.

\subsection{Exceptional points}

{In Hermitian two-band Floquet systems, the closing of two bands occurs at
either $\varepsilon \left( \mathbf{k}_{c}\right) =0$ or $\varepsilon \left(
\mathbf{k}_{c}\right) =\pi $ \cite{Longwen}. In the presence of
non-Hermiticity here, i.e., $\mathbf{g}_{i}\left( \mathbf{k}\right) \neq 0$,
the Floquet band closing points are in fact exceptional points where two
Floquet eigenstates coalesce. This can be seen from the behavior of Floquet
operator $U\left( \mathbf{k}\right) =\cos \left[ \varepsilon \left( \mathbf{k%
}\right) \right] -i\sin \left[ \varepsilon \left( \mathbf{k}\right) \right]
\widehat{\mathbf{n}}\left( \mathbf{k}\right) \cdot \mathbf{\sigma }$ at the
band touching points {$\varepsilon \left( \mathbf{k}_{c}\right) =0$ or $%
\varepsilon \left( \mathbf{k}_{c}\right) =\pi $,} which yields the identity $%
\cos \left[ \varepsilon \left( \mathbf{k}_{c}\right) \right] =\pm 1$.
However, the presence of non-Hermitian term {$\mathbf{g}_{i}\left( \mathbf{k}%
_{c}\right) $ does not need }$\mathbf{n}\left( \mathbf{k}_{c}\right) =0$ but
leads to $\sin \left[ \varepsilon \left( \mathbf{k}_{c}\right) \right]
=\left\Vert \mathbf{n}\left( \mathbf{k}_{c}\right) \right\Vert =0$. Hence
the Floquet operator at the band closing points reduces to
\begin{equation}
U\left( \mathbf{k}_{c}\right) =\pm 1-i\mathbf{n}\left( \mathbf{k}_{c}\right)
\cdot \mathbf{\sigma ,}
\end{equation}%
with $\left\Vert \mathbf{n}\left( \mathbf{k}_{c}\right) \right\Vert =0$. It
turns out that for our model systems below, $U\left( \mathbf{k}_{c}\right) $
obtained above is a Jordan block form accompanied by the coalescence of two
eigenstates $\left\{ \left\vert \varphi \right\rangle \right\} $. For
example. Suppose $\mathbf{n}\left( \mathbf{k}_{c}\right) =\left( 1\text{, }0%
\text{, }i\right) $ such that $\mathbf{n}\left( \mathbf{k}_{c}\right) \neq 0$
but $\left\Vert \mathbf{n}\left( \mathbf{k}_{c}\right) \right\Vert =0$, the
associated Floquet operator then becomes
\begin{equation}
U\left( \mathbf{k}_{c}\right) =\pm 1-i\left(
\begin{array}{cc}
i & 1 \\
1 & -i%
\end{array}%
\right) .
\end{equation}%
The two coalescent eigenstates are $\left\vert \varphi \right\rangle =\left(
i,\text{ }1\right) ^{T}$. In this sense, the Floquet operator cannot be
diagonalized as it can be transformed to the Jordan block form through the
similarity transformation $V^{-1}\left( \mathbf{n}\left( \mathbf{k}%
_{c}\right) \cdot \mathbf{\sigma }\right) V$, where $V=[v_{1}$, $v_{2}]$
with $v_{1}=\left( i,\text{ }1\right) ^{T}$ and $v_{2}=\left( 1,\text{ }%
0\right) ^{T}$. Thus, the two bands touch at EPs, which is markedly
different from what is known in Hermitian Floquet system. Indeed, a DP at $%
\mathbf{k}=\mathbf{k}_{c}$ in a Hermitian Floquet two-band system would
requires $\mathbf{n}\left( \mathbf{k}_{c}\right) =0$ and $\left\Vert \mathbf{%
n}\left( \mathbf{k}_{c}\right) \right\Vert =0$. Given that DPs can be deemed
as topological defects in the quasi-energy spectrum in Hermitian systems, it
would be curious to examine how DPs turn themselves into EPs as we gradually
introduce non-Hermiticity into the system. Previously it was shown (in
static systems however) that a DP can split into two EPs \cite{Lee}. In our
specific examples elaborated below, we shall inspect if this splitting also
occurs, and if yes, whether each of the two EPs can also \textquotedblleft
inherit" half topological invariant of the parent DPs. 
}

\subsection{Floquet non-Hermitian skin effect}

Now we turn to another well-known non-Hermitian effect, i.e. NHSE. Due to
NHSE, there is a marked difference between band structure under OBC and that
under PBC \cite{Chinghua}. {Generically, OBC skin modes can be extrapolated
from the PBC eigenmodes upon inserting a complex flux to the system with
increasing amplitude. During that process, the PBC spectra, which are
generically closed loops in the complex plane as a function of
quasi-momentum, collapse into open lines or arcs that precisely represent
the edge modes under OBCs \cite{Chinghua}. It is hence curious to examine
NHSE in periodically driven systems. Interestingly, to our knowledge, except
for a quantum walk experimental study \cite{Xuepeng2019}, little is known
about Floquet NHSE (FNHSE). Following \cite{Chinghua,Yao1,Yokomizo}, it is
now understood that if for each $\mathbf{k}$, there always exists $\mathbf{k}%
^{\prime }$ such that $\varepsilon \left( \mathbf{k}\right)= \varepsilon
\left( \mathbf{k}^{\prime }\right)$, then the PBC spectrum does not possess
any loop structure in the first place and hence FNHSE cannot present under
OBC. On the other hand, if such a relation does not hold, then the spectrum
may form some loop structure in the complex quasi-energy plane.  As a result, FNHSE
can be manifested drastically upon introducing OBC.}

\subsection{Winding numbers}

{In this subsection, we will define two different types of winding numbers
to capture the topological property of the Floquet spectrum uner PBC. As
seen below, the first type of winding numbers reflects the vorticity of the
quasi-energy spectrum and the second type of winding numbers characterize
the BBC instead. Together they offer complementary topological
characterization of the model systems we shall introduce later.}

Consider first two specific time windows from $t=T_{1}/2$ to $%
t=3T_{1}/2+T_{2}$ and $t=T_{2}/2$ to $t=3T_{2}/2+T_{1}$. The second time
frame represents a shift of first time frame. The resulting respective
Floquet operators in such two symmetric time frames are given by%
\begin{equation}
U_{i}\left( \mathbf{k}\right) =n_{0}+i\mathbf{n}_{i}\left( \mathbf{k}\right)
\cdot \mathbf{\sigma }\text{, }i=1,\text{ }2  \label{U_timeframe}
\end{equation}%
where%
\begin{eqnarray}
n_{0} &=&\cos \left[ \varepsilon \left( \mathbf{k}\right) \right] , \\
\mathbf{n}_{i}\left( \mathbf{k}\right) &=&[\cos \left( \left\vert \mathbf{B}%
_{i}\right\vert T_{i}\right) \sin \left( \left\vert \mathbf{B}%
_{j}\right\vert T_{j}\right) \widehat{\mathbf{B}}_{i}\left( \mathbf{k}%
\right) \cdot \widehat{\mathbf{B}}_{j}\left( \mathbf{k}\right)  \notag \\
&&\left. +\sin \left( \left\vert \mathbf{B}_{i}\right\vert T_{i}\right) \cos
\left( \left\vert \mathbf{B}_{j}\right\vert T_{j}\right) \right] \cdot
\widehat{\mathbf{B}}_{i}\left( \mathbf{k}\right)  \notag \\
&&+\sin \left( \left\vert \mathbf{B}_{j}\right\vert T_{j}\right) [\widehat{%
\mathbf{B}}_{j}\left( \mathbf{k}\right)  \notag \\
&&-(\widehat{\mathbf{B}}_{i}\left( \mathbf{k}\right) \cdot \widehat{\mathbf{B%
}}_{j}\left( \mathbf{k}\right) )\widehat{\mathbf{B}}_{i}\left( \mathbf{k}%
\right) ],
\end{eqnarray}%
with $i=1$, $j=2$ or $i=2$, $j=1$. To digest the results here, we consider a
periodically quenched system with sublattice symmetry, i.e., $%
SH_{i}S^{-1}=-H_{i},$ where $S$ represents certain chiral-symmetry (unitary)
operators. Because there is no cross product of any two Pauli matrices
appearing the expressions above, it can be easily seen that the Floquet
operator still possesses the same chiral symmetry, namely, $SU_{i}\left(
\mathbf{k}\right) S^{-1}=U_{i}^{-1}\left( \mathbf{k}\right) $. As such, the
effective Floquet Hamiltonian $H_{\mathrm{eff}}^{j}\left( \mathbf{k}\right)
=i\ln U_{j}\left( \mathbf{k}\right) /T$ with $j=1,2$ also contains only
two Pauli matrices such that the complex-quasi energy spectrum is symmetric
with respect to zero.

As an example, we assume that $H_{\mathrm{eff}}^{j}\left( \mathbf{k}\right)$
only includes two Pauli matrices and commutes with $\sigma _{y}$. That is,
\begin{eqnarray}  \label{equin}
H_{\mathrm{eff}}^{j}\left( \mathbf{k}\right)= n_{j,x} \left( \mathbf{k}%
\right) \sigma_{x} +n_{j,z} \left( \mathbf{k}\right) \sigma _{z},
\end{eqnarray}
with with $(j=1,2)$. In the Hermitian case, one then uses the winding of the
vector $[(n_{j,x} \left( \mathbf{k})\right), n_{j,z} \left( \mathbf{k}%
\right)]$ as a function of $\mathbf{k}$ in the $xz$-plane to define winding
numbers, and this vector would be ill-defined at any DP because $[(n_{j,x}
\left( \mathbf{k})\right), n_{j,z} \left( \mathbf{k}\right)]=0 $ for a DP.
However, at EPs in non-Hermitian systems, as illuminated above, the $%
[(n_{j,x} \left( \mathbf{k})\right), n_{j,z} \left( \mathbf{k}\right)]$ is
generally nonzero. So we need to resort an alternative definition of winding
numbers. This is made possible by the observation that at any EP at zero or $%
\pi$ quasi-energy, the expectation value of $H_{\mathrm{eff}}^{j}\left(
\mathbf{k}\right)$ on the associated quasi-energy eigenstate must be zero.
As a result, the corresponding expectation values of $\sigma_{x}$ and $%
\sigma_{x}$ must be zero at such EPs. This facilitates us to define a vector
field
\begin{equation*}
\mathbf{D}_{j}\left( \mathbf{k,s}\right) =\left( \left\langle \sigma
_{x}\right\rangle _{j},\text{ }\left\langle \sigma _{z}\right\rangle
_{j}\right),
\end{equation*}%
where $\mathbf{s}$ denotes other system parameters. This vector field
becomes ill-defined at an EP, indicating that EPs represent topological
defects under this new definition of winding numbers for non-Hermitian
systems \cite{Sunk,Hou}. 

In a more formal language, the topological characterization of the effective
Floquet Hamiltonian of $H_{\mathrm{eff}}^{j}\left( \mathbf{k}\right) $ can
be captured by{%
\begin{equation}
w_{\mathrm{I},j}=\frac{1}{2\pi i}\int\nolimits_{C}\Xi _{j,S}^{-1}\text{ }%
\mathrm{d}\Xi _{j,S},\text{ }j=1,\text{ }2  \label{w1}
\end{equation}%
}where $\Xi _{j,S}=\left\langle \sigma _{x}\right\rangle _{j}+i\left\langle
\sigma _{z}\right\rangle _{j}$ and $C$ denotes a closed loop in the $%
\left\langle \sigma _{x}\right\rangle _{j}-\left\langle \sigma
_{z}\right\rangle _{j}$ vector field space. This closed loop can be
specified after we define an arbitrary closed loop in the momentum space
first and then accordingly compute the expectation values $\left\langle
\sigma _{x}\right\rangle _{j}$ and $\left\langle \sigma _{z}\right\rangle
_{j}$ on the associated Floquet eigenstates as a function of $\mathbf{k}$.
As elaborated below, choosing different closed loops in the momentum space
will result in different closed loops in the vector field space and hence
different winding numbers. The sign of $w_{\mathrm{I},j}$ also indicates the
winding direction of vector field $\mathbf{D}_{j}\left( \mathbf{k,s}\right) $%
. Since Floquet operators associated with the two different time-symmetry
time frames are connected by a unitary transformation, which does not change
the location of EPs, the two Floquet effective Hamiltonians in both
symmetric time frames should yield equivalent topological winding numbers.
Here we want to stress that EPs play the same role as DPs in the parent
Hermitian Floquet system. The value of the winding number depends on whether
the loop in the momentum space encircle the EP. In the following concrete
examples, the different topological phases can be identified with the aid of
$w_{\mathrm{I},j}$, especially in the 1D example. In this sense, the EP can
be deemed as the topological phase transition points.
\begin{figure}[tbp]
\centering
\includegraphics[width=0.45\textwidth]{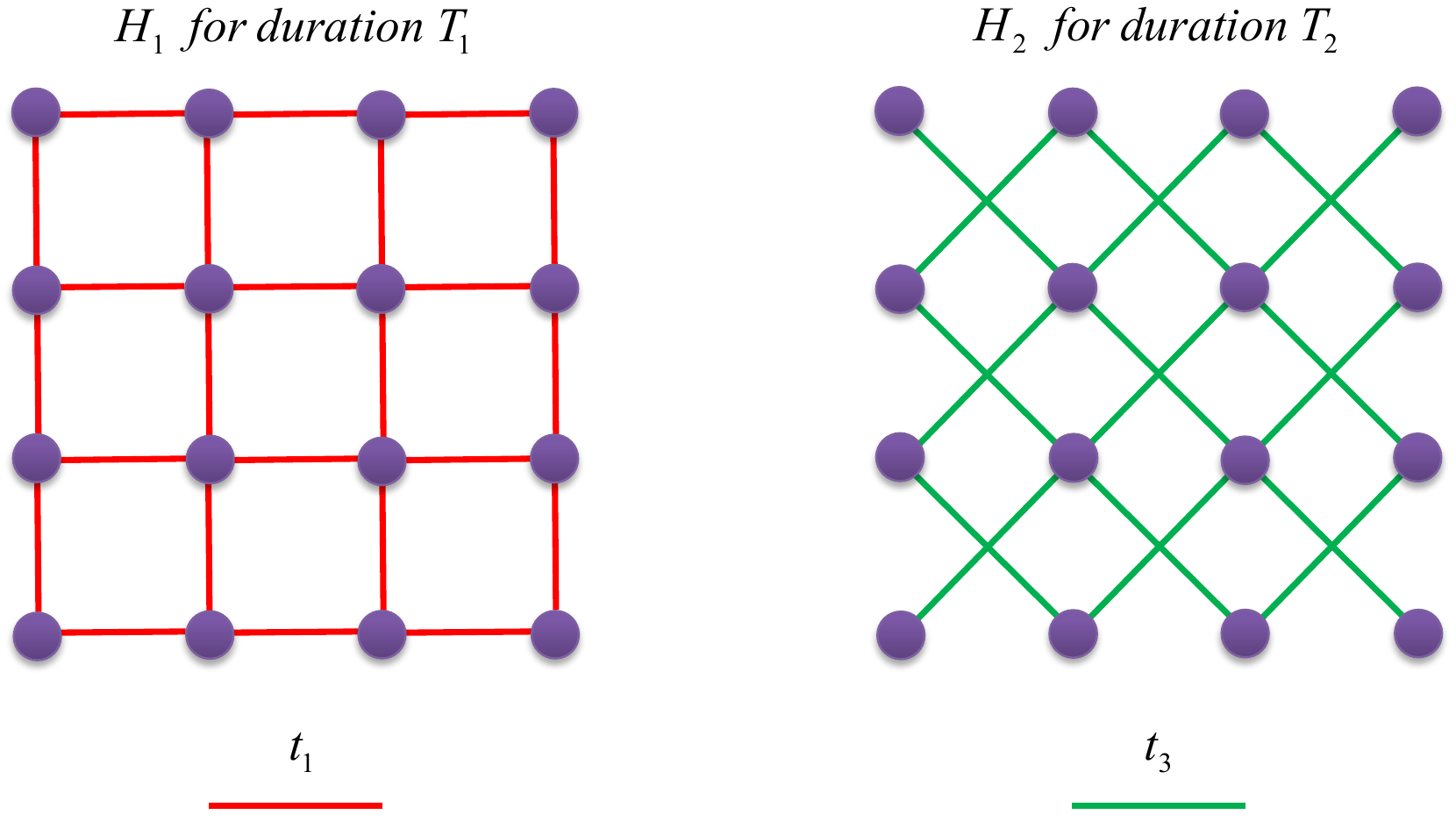}
\caption{Schematic illustration of a $2D$ quenched system. Each purple dot
represents a unit cell with two internal degrees of freedom. The system
Hamiltonian is $H_{1}$ for duration $T_{1}$ for which the couplings between
the adjacent unit cells are only along the square-side directions. The
system Hamiltonian is then quenched to $H_{2}$ for duration $T_2$, where
only intercell coupling along the diagonal direction exists.}
\label{figure_illu_2D}
\end{figure}

Next we turn to the second definition of winding numbers $w_{\mathrm{II},j}$%
, analogous to some known treatments in Hermitian cases. That is, {%
\begin{equation}
w_{\mathrm{II},j }=\frac{1}{2\pi i}\int\nolimits_{C_{\mathbf{k}}}\mathrm{d}%
\mathbf{k}\text{ }L_{j,S}^{-1}\left( \mathbf{k}\right) \frac{\mathrm{d}}{%
\mathrm{d}\mathbf{k}}L_{j,S}\left( \mathbf{k}\right) ,  \label{w2}
\end{equation}%
}where $L_{j,S}\left( \mathbf{k}\right) =n_{j, x}\left( \mathbf{k}\right)
+in_{j,z}\left( \mathbf{k}\right) $, with $n_{j,x}\left( \mathbf{k}\right)$
and $n_{j,z}\left( \mathbf{k}\right)$ still describing the effective Floquet
Hamiltonian [see Eq.~(\ref{equin})]. {At first glance this definition seems
to advocate complex winding numbers. However, due to the assumed chiral
symmetry of the system, it can be shown that the winding numbers thus
defined are actually real \cite{Longwen}.} Note also that $w_{\mathrm{II},j}$
represents and only represents the winding behavior of $L_{j,S}\left(
\mathbf{k}\right)$ around the origin of the momentum space (this is thus
different from our choice of a rather arbitrary closed loop in the momentum space when
examining the first type of winding numbers). Following previous extensive
results in Hermitian Floquet systems, winding numbers $w_{\mathrm{II},j}$ in
two time-symmetric frames are expected to predict the number of Floquet edge
states, thereby reflecting BBC. Specifically, using $w_{\mathrm{II},1}$ and $%
w_{\mathrm{II},2}$, one can obtain $W_{0}$ and $W_{\pi }$ as {%
\begin{equation}
W_{0}=\frac{w_{\mathrm{II},1}+w_{\mathrm{II},2}}{2}\text{, }W_{\pi }=\frac{%
w_{\mathrm{II},1}-w_{\mathrm{II},2}}{2}.  \label{zero_and_pi}
\end{equation}%
}which predict the number of edge state pairs at quasi energies $\varepsilon
\left( \mathbf{k}\right) =0$ and $\varepsilon \left( \mathbf{k}\right) =\pi $
\cite{Longwen}. Importantly, this BBC perspective will be challenged by
FNHSE. Indeed, in the presence of FNHSE, such bulk topological invariants
can no longer predict the emergence of topological edge modes because
non-reciprocity in non-Hermitian systems can fundamentally alter the
non-Bloch quasi-energy spectrum \cite{Yao1,Yokomizo}. It is hence curious to
understand how to recover BBC in non-Hermitian Floquet systems with FNHSE.
As seen below via one concrete example, we must introduce the concept of the
so-called generalized Brillouin zone $C_{\mathbf{\beta }}$ to replace $C_{%
\mathbf{k}}$ in Eq.~(\ref{w2}), so as to recalculate the second-type winding
numbers.

\section{Model: 2D non-Hermitian Floquet system}

\begin{figure}[tbp]
\centering
\includegraphics[width=0.4\textwidth]{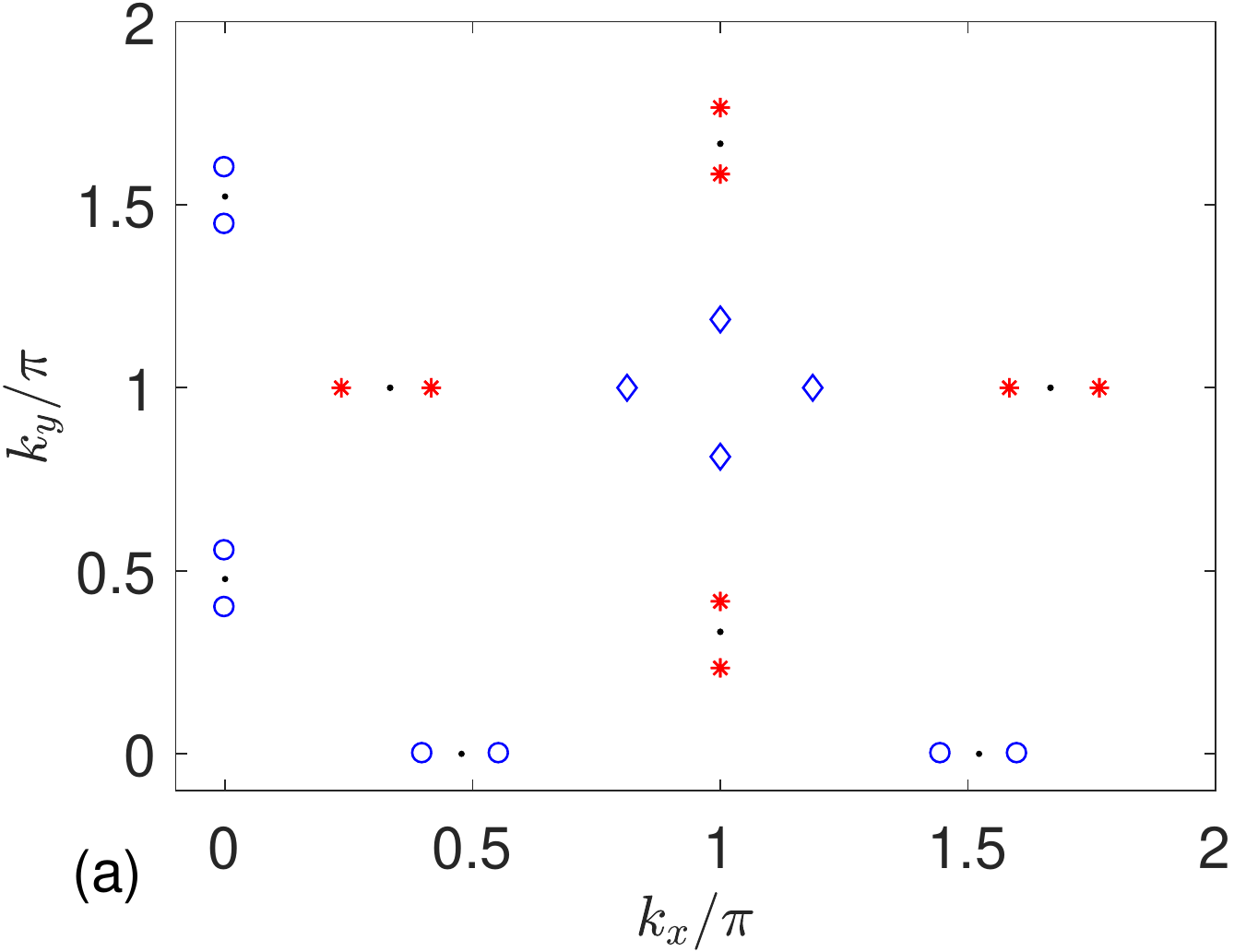} %
\includegraphics[width=0.4\textwidth]{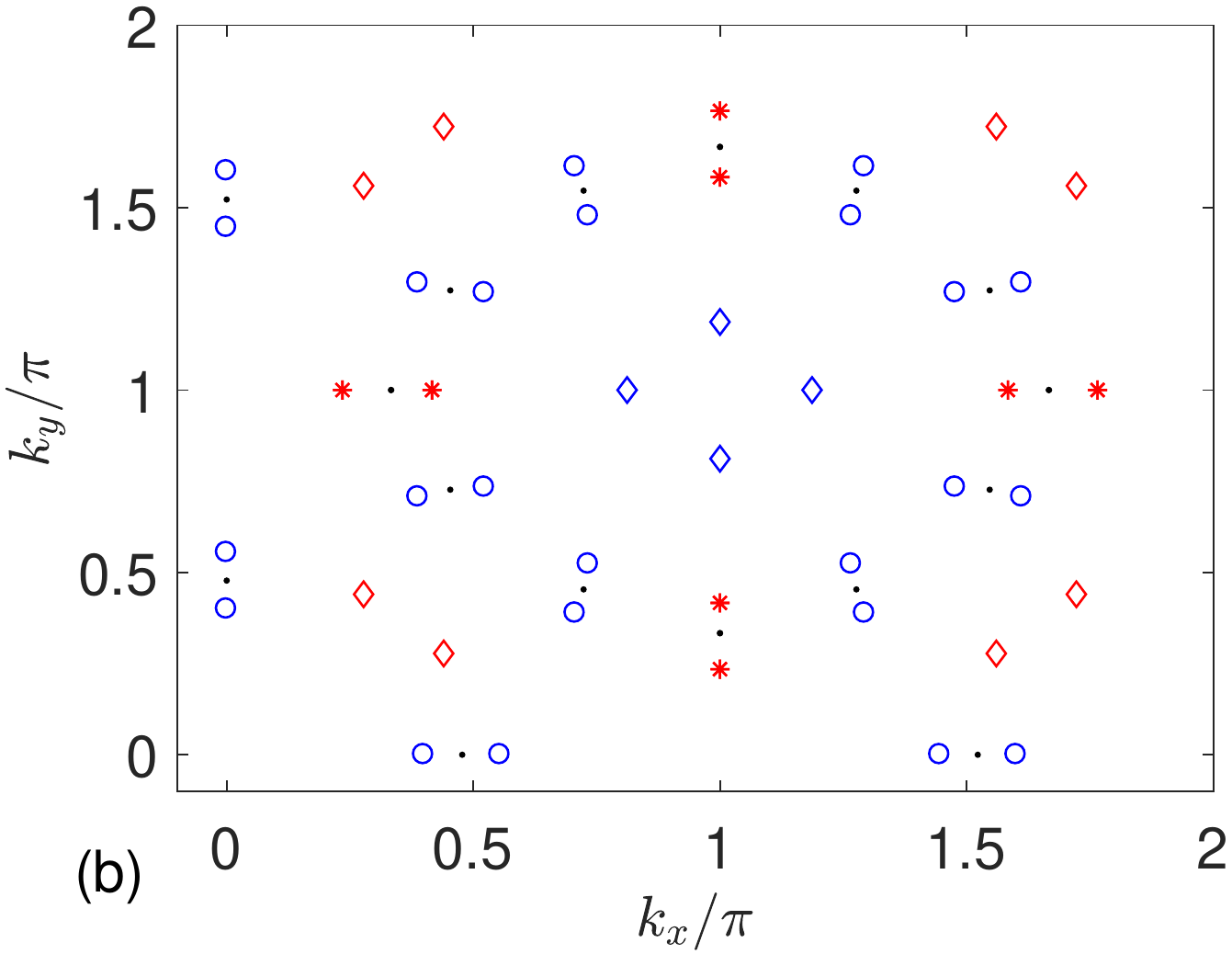}
\caption{Many EPs in the PBC spectrum of a 2D Floquet system illustrated in
Fig. 1. Detailed descriptions of this model are specified in the main text,
with (a) $t_{1}=1$, $\protect\gamma =0.5$, $t_{2}=1$ and $t_{3}=0.5$. (b) $%
t_{1}=1$, $\protect\gamma =0.5$, $t_{2}=1$ and $t_{3}=\protect\pi /3$. The
location of the EPs are determined by Eqs. (\protect\ref{EP_2D1})-(\protect
\ref{EP_2D2}). The different parities yield different quasi-energies of the
EPs, which are denoted by red ($0$) and blue labels ($\protect\pi $),
respectively. Black points represent the DPs of the parent Hermitian Floquet
system without the non-Hermitian term ($\protect\gamma =0$). Red and blue
diamonds denote type-II EPs at $0$ and $\protect\pi $ quasi-energies (that
is, not resulting from the splitting of the DPs of the parent Hermitian
system). It is seem that each of the DPs split into two EPs due to
non-Hermiticity and therefore each of the EPs inherits half topological
winding numbers of the parent DPs.}
\label{fig_split}
\end{figure}

\begin{figure*}[tbp]
\centering
\includegraphics[width=0.42\textwidth]{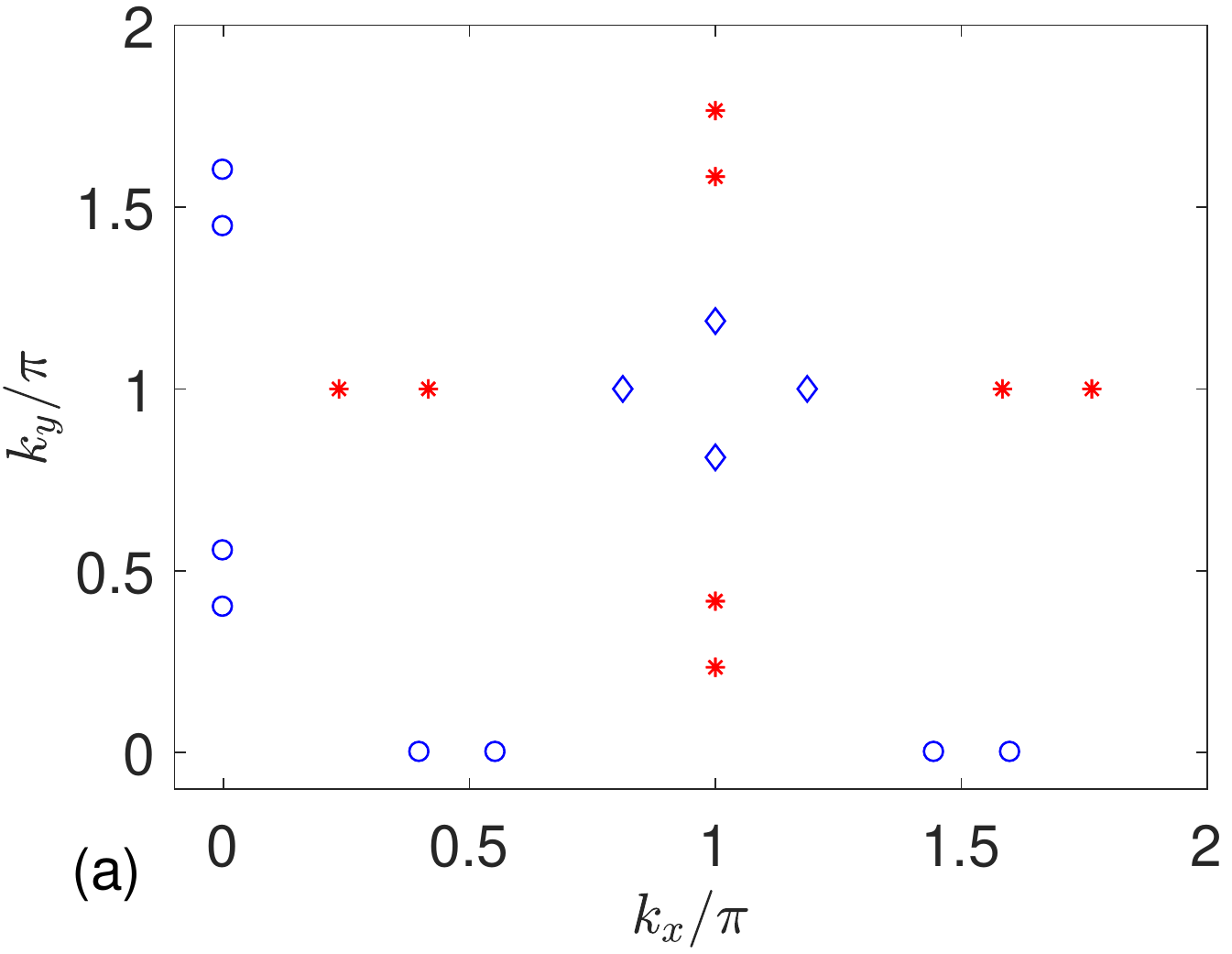} %
\includegraphics[width=0.42\textwidth]{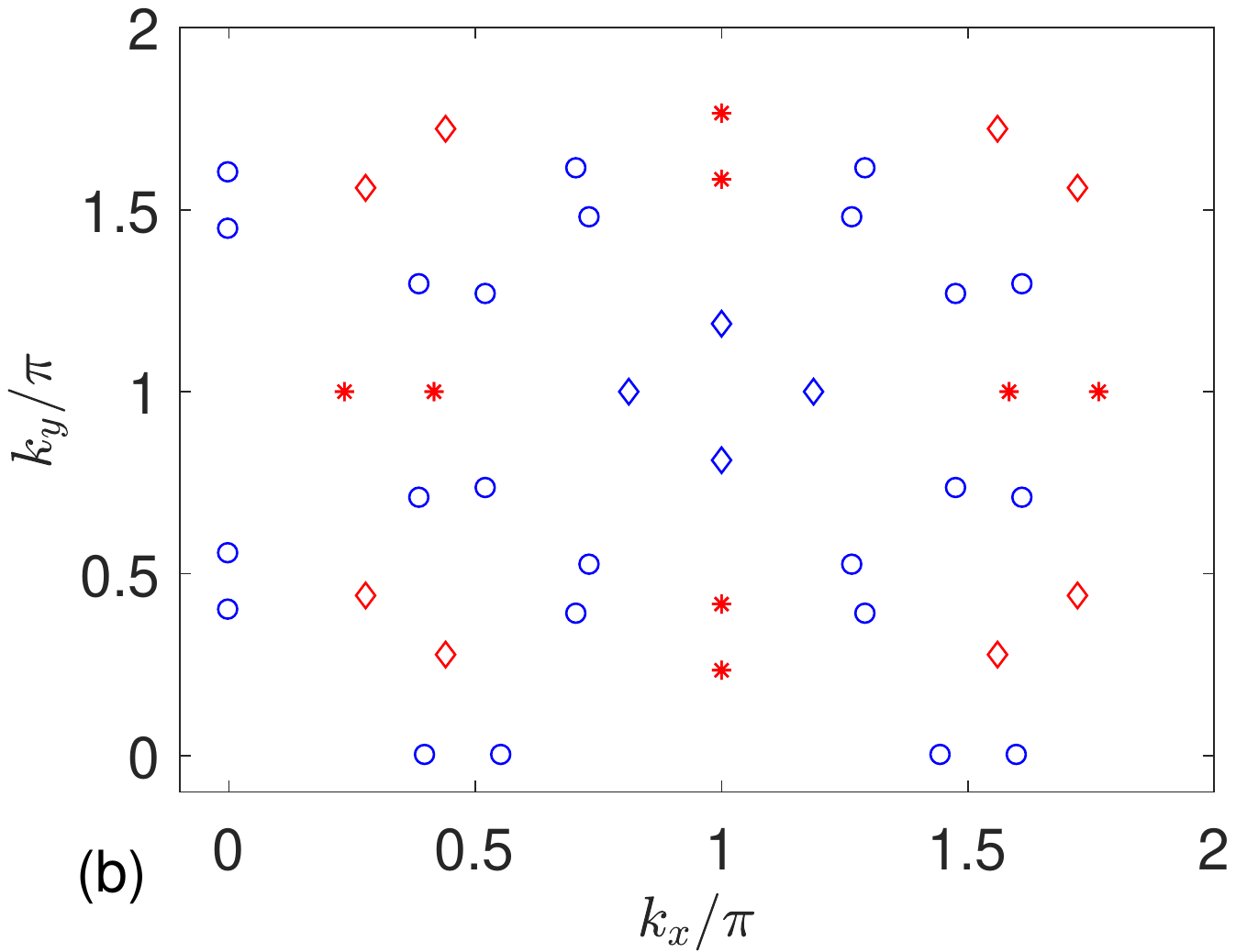} %
\includegraphics[width=0.42\textwidth]{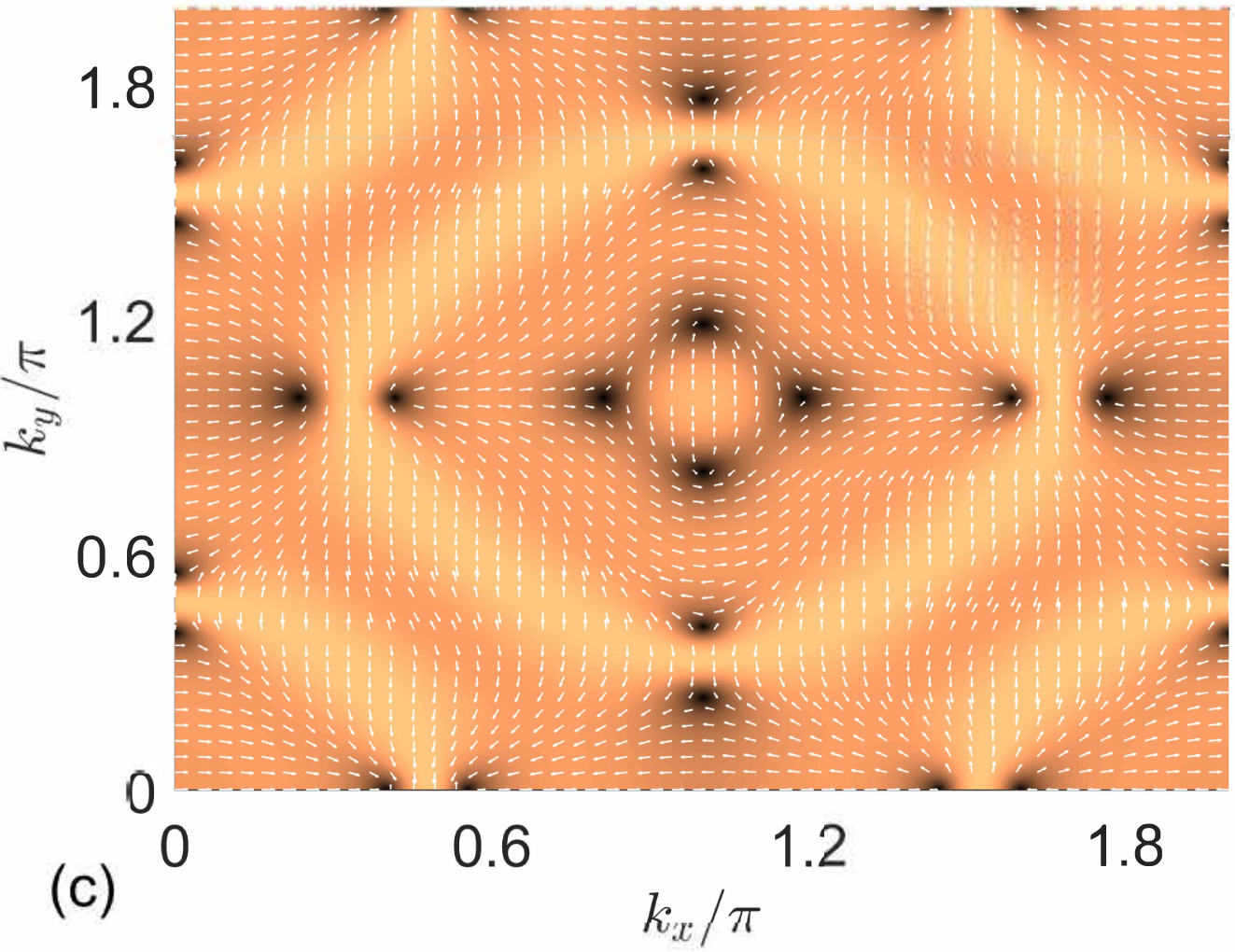} %
\includegraphics[width=0.42\textwidth]{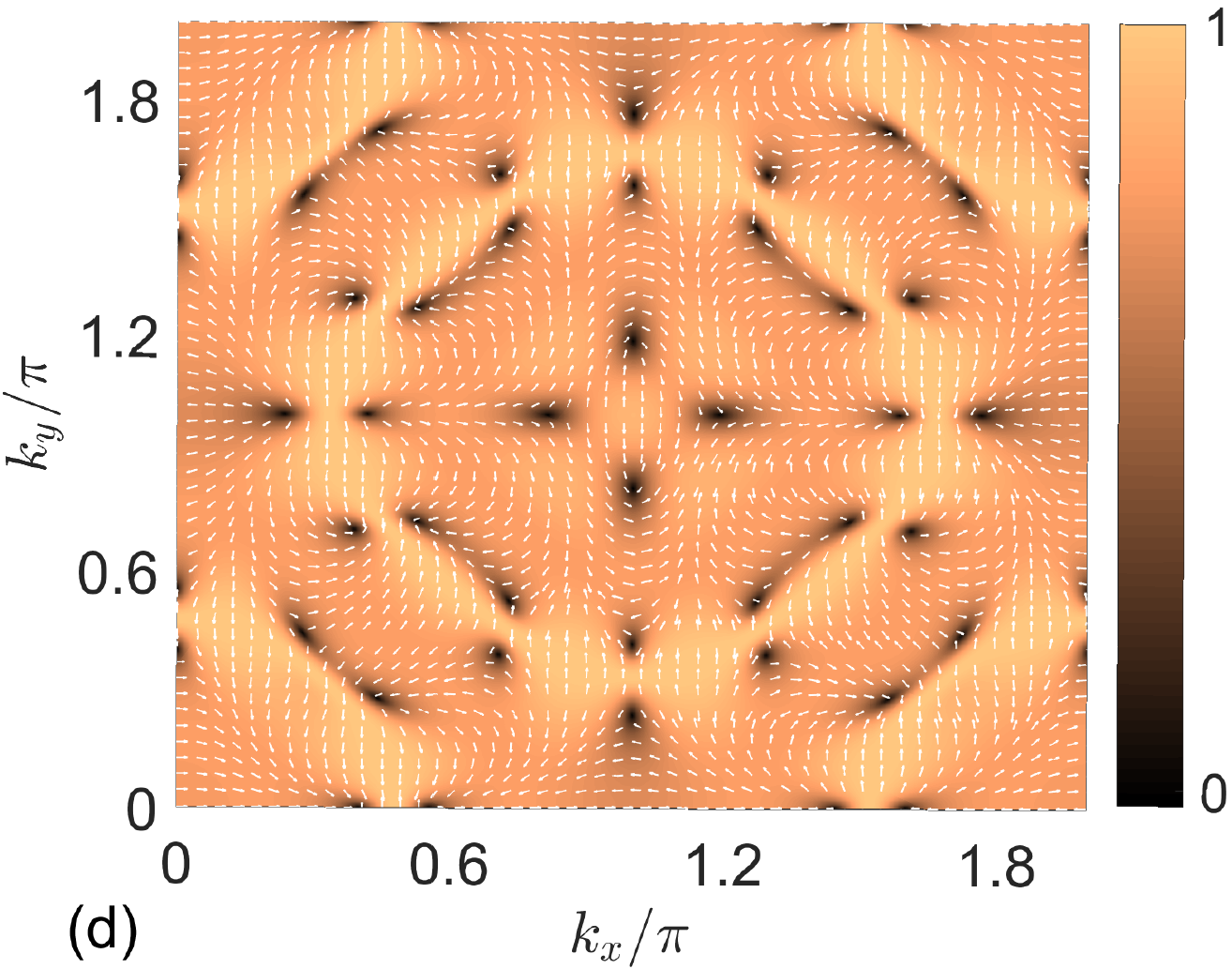}
\caption{Plots of the planar vector field $\mathbf{D}_{1}\left( \mathbf{k}%
\right) $ and EPs as defined in the main text. (a)-(b) The EPs with
quasi-energies $0$ and $\protect\pi $ denoted by blue and red circles with
the same system parameters as in Fig. \protect\ref{fig_split}. (c-d) The
corresponding planar vector fields. The density of the background plot is $%
\left( \left\langle \protect\sigma _{x}\right\rangle _{1}\right) ^{2}+\left(
\left\langle \protect\sigma _{z}\right\rangle _{1}\right) ^{2}$. The black
regions show the location of EPs with the condition of $\left\langle \protect%
\sigma _{x}\right\rangle _{1}=\left\langle \protect\sigma _{z}\right\rangle
_{1}=0$. When the vector field surrounds an EP, its direction will only
change angle $\protect\pi$ (hence winding number is $\pm 1/2$), which is not
possible in Hermitian systems. This will be also demonstrated in Fig.~%
\protect\ref{fig_trajectory}.}
\label{fig_vector}
\end{figure*}

\begin{figure}[tbp]
\centering
\includegraphics[width=0.42\textwidth]{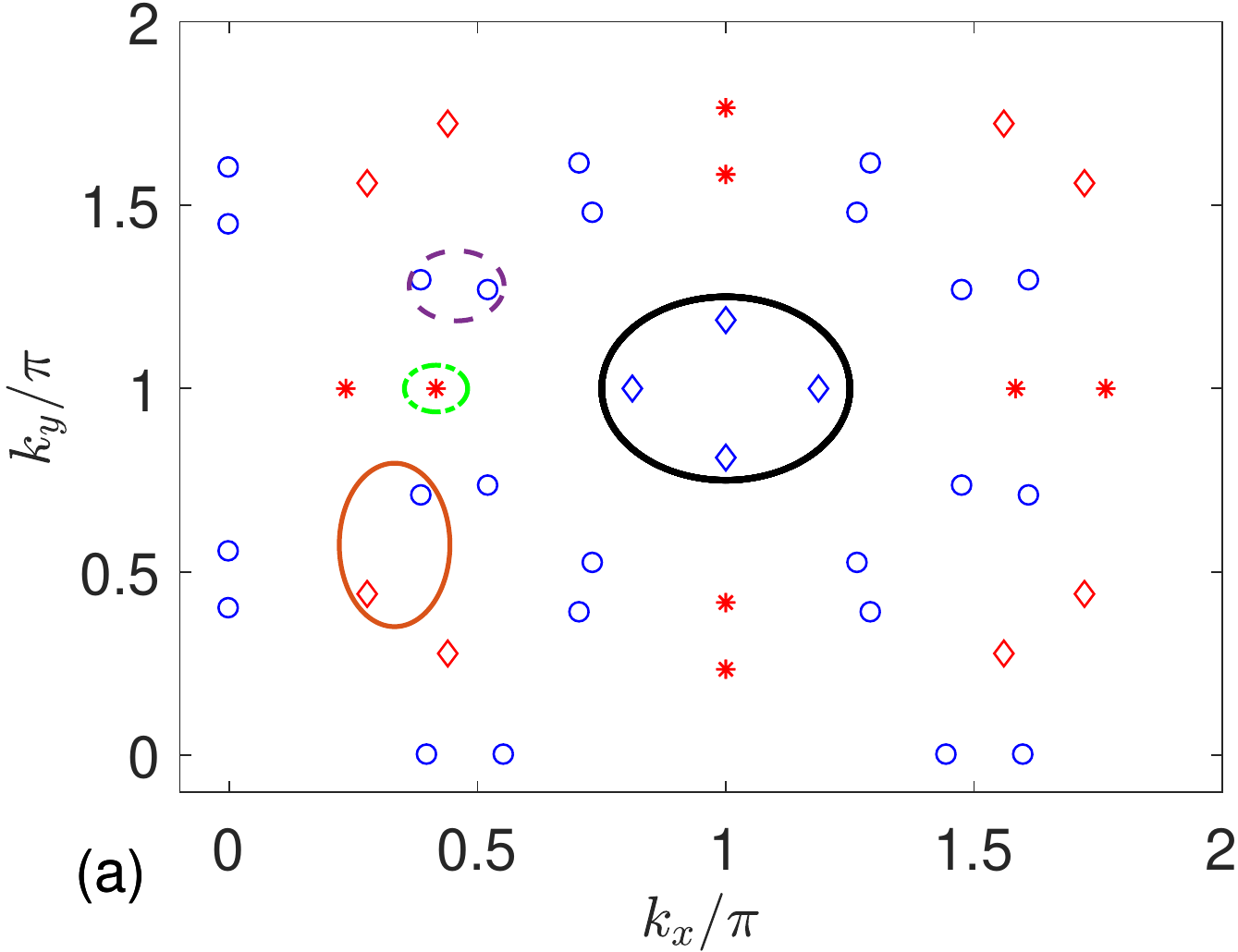} %
\includegraphics[width=0.42\textwidth]{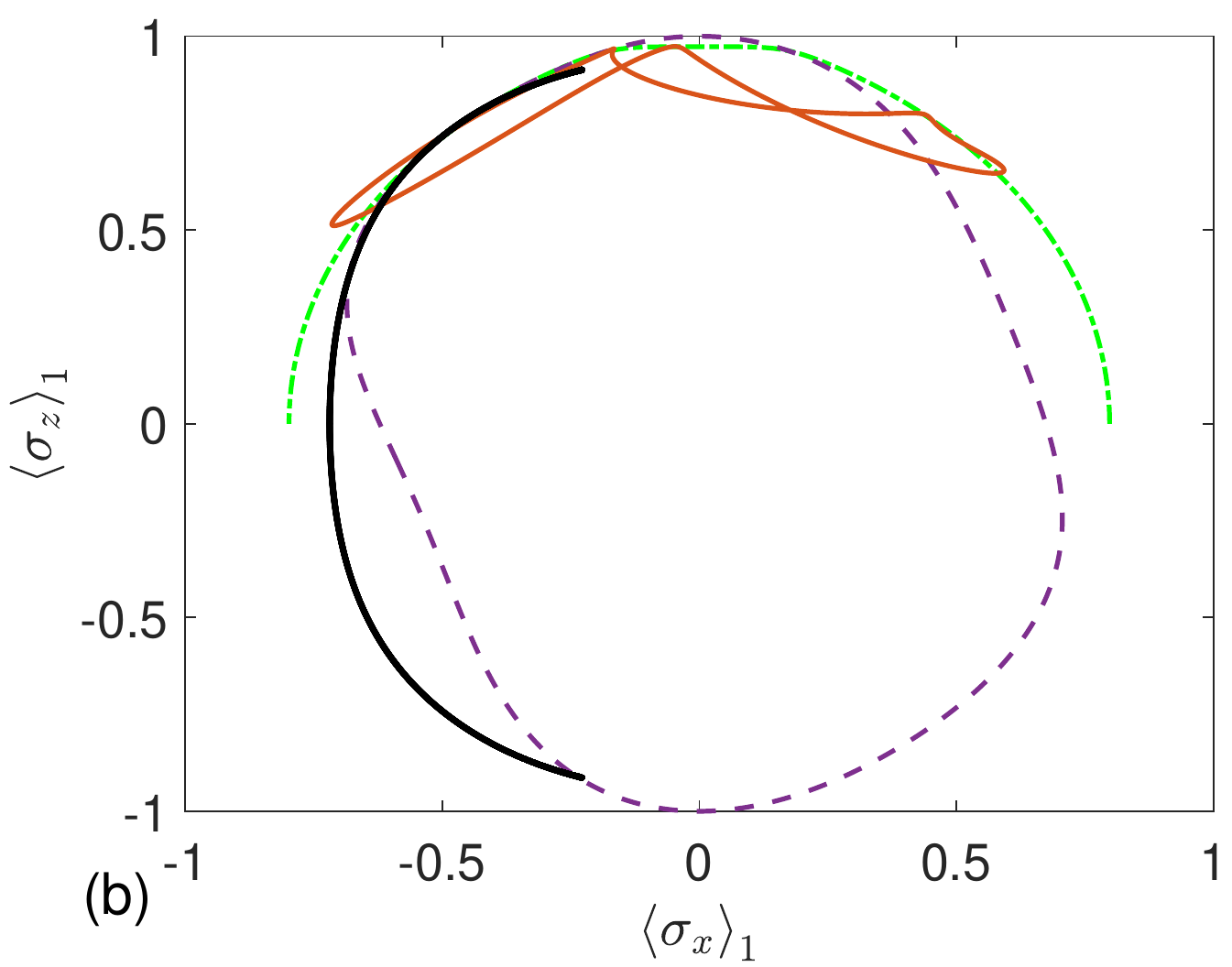}
\caption{Trajectory of ($\left\langle \protect\sigma _{x}\right\rangle _{1}$%
, $\left\langle \protect\sigma _{z}\right\rangle _{1})$ obtained along a
momentum-space closed loop. Green, purple, orange, and black lines are for
results when a closed loop encircles an EP, two EPs with the same
quasi-energy, two EPs with different quasi-energies and four type-II EPs.
The system parameters are the same with those in \protect\ref{fig_split}
(b). It is seen that when the loop surrounds one EP, the vector field $%
\mathbf{D}_{1}\left( \mathbf{k}\right) $ does wind around the origin, but
the momentum-space loop used to define the trajectory must continue to wind
twice in order to form a closed trajectory. Hence the winding number $%
w_{I,1} $ has a fractional value $\pm 1/2$. When the momentum-space loop
encircles two EPs with same (opposite) charge, the obtained winding number
will be $1$ (0). Note that the points of black curve are double degenerate
and hence the winding number is $0$.}
\label{fig_trajectory}
\end{figure}

\label{two_examples} With all the preparations above, we are now ready to
develop more physical insights by working on a concrete system.

Let us consider a 2D non-Hermitian Floquet system, the two Hamiltonians that
the system is periodically quenched between are assumed to be
\begin{eqnarray}
H_{1} &=&\sum_{k_{x},k_{y}}B_{1,x}\left( k_{x},k_{y}\right) \sigma
_{x}\left\vert k_{x},k_{y}\right\rangle \left\langle k_{x},k_{y}\right\vert ,
\label{H1_2D} \\
H_{2} &=&\sum_{k_{x},k_{y}}B_{2,z}\left( k_{x},k_{y}\right) \sigma
_{z}\left\vert k_{x},k_{y}\right\rangle \left\langle k_{x},k_{y}\right\vert ,
\label{H2_2D}
\end{eqnarray}%
where
\begin{eqnarray}
B_{1,x}\left( k_{x},k_{y}\right) &=&2t_{1}\left( \cos k_{x}+\cos
k_{y}\right) +t_{2}, \\
B_{2,z}\left( k_{x},k_{y}\right) &=&4t_{3}\sin k_{x}\sin k_{y}+i\gamma .
\end{eqnarray}%
with $t_{1}$ and $t_{3}$ being the intercell couplings along the square-side
and diagonal directions. $t_{2}$ describes the intracell coupling and $%
\gamma $ represents the balance gain and loss in each unitcell. Obviously,
the Floquet operators (\ref{U_timeframe}) of two symmetric time frames
respect the chiral symmetry, i.e., $\sigma _{y}U_{i}\left(
k_{x},k_{y}\right) \sigma _{y}^{-1}=U_{i}^{-1}\left( k_{x},k_{y}\right) $.
In Fig. \ref{figure_illu_2D}, we sketch two alternating lattice
configurations of the system. In addition to the rather familiar optical
lattice and waveguide array platforms, such 2D non-Hermitian Floquet system
could be also realized with circuits and nonreciprocal electrical elements
or in discrete-time non-unitary quantum-walk system \cite{Helbig,Xuepeng2019,Ghatak,Hofmann}. Note also that in the high-frequency limit $%
T\rightarrow 0$ (with $T_1=T_2$), the Floquet effective Hamiltonian $H_{\mathrm{eff}}^{i}$ is
simply given by $(H_1+H_2)/2$, which in the coordinate space represents a
tight-binding bilayer square lattice \cite{ZhangKL}. In that limit the EPs
always occur at either $k_{x}=0$, $\pi $ or $k_{y}=0$, $\pi $. Thus, it is
interesting to see how periodic quenching in general may change the
locations of the EPs.
\begin{figure*}[tbp]
\centering
\includegraphics[width=0.42\textwidth]{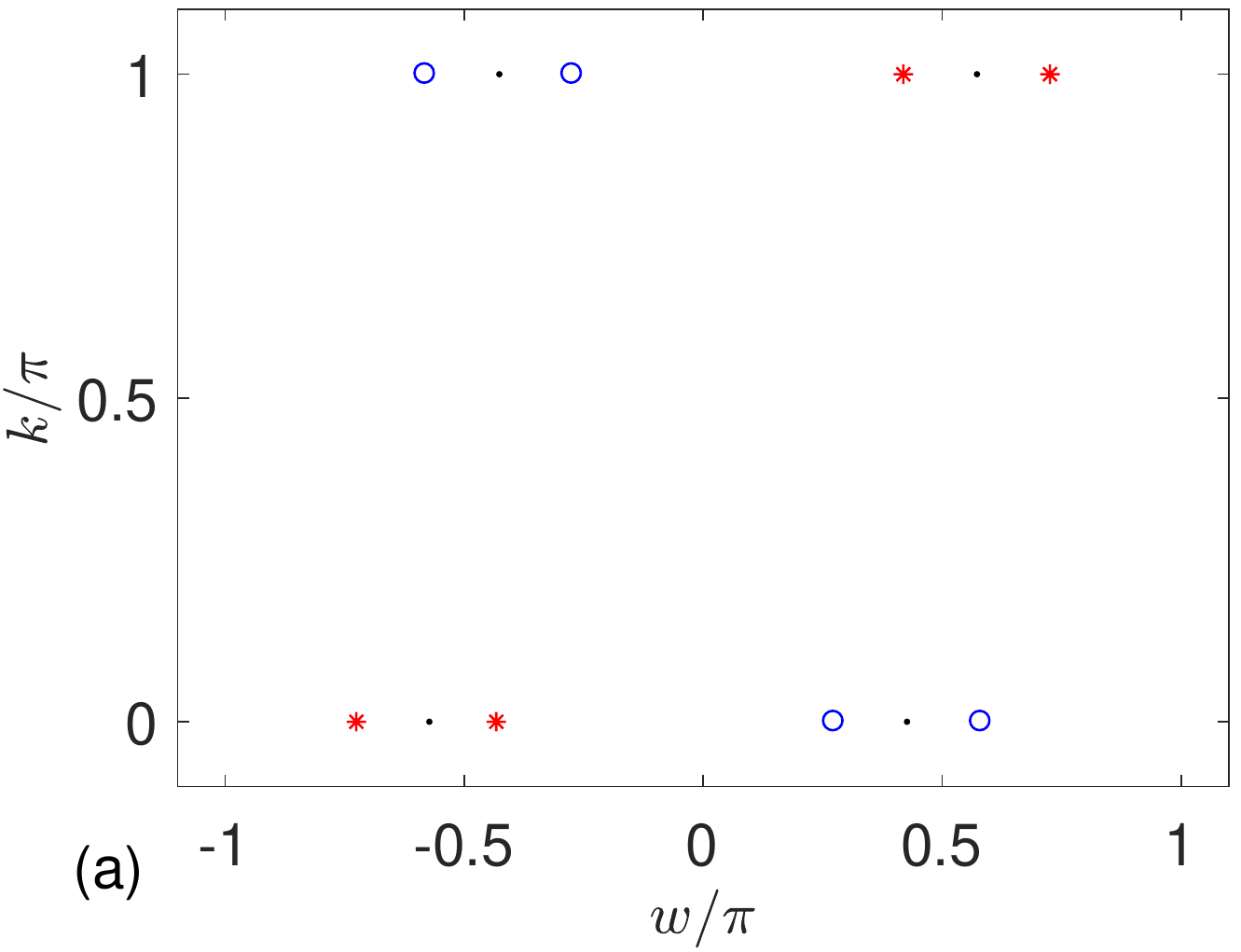} %
\includegraphics[width=0.42\textwidth]{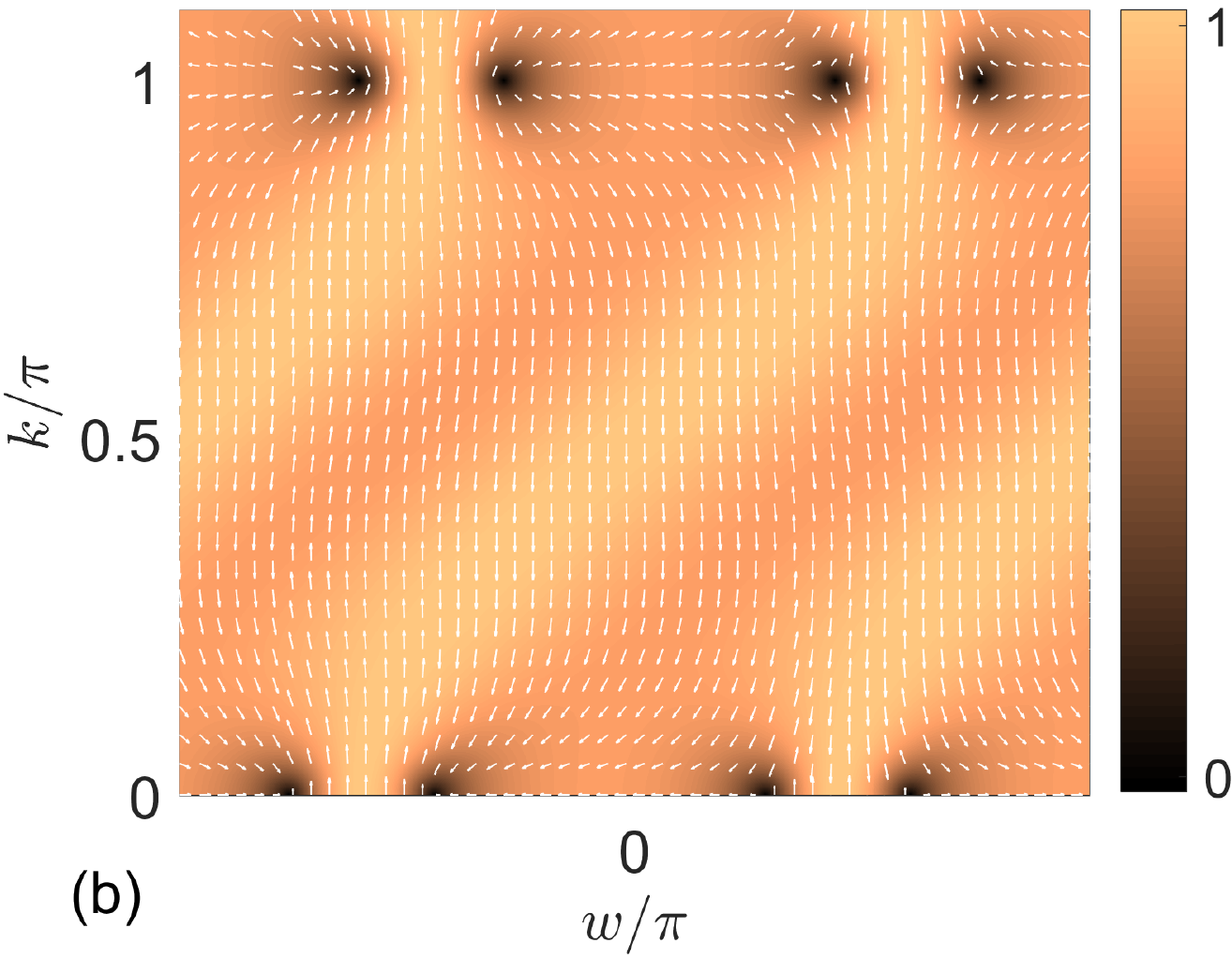}
\caption{Many EPs found in our 1D non-Hermitian Floquet system. Each of the
EPs carries half-integer topological defect, as manifested by the planar
vector field $\mathbf{D}_{i}\left(k,w\right) $. The system parameters are $%
v=1.8$, $\protect\gamma =0.5$. Note that there is no type-II EPs for 1D
model here. All the EPs stem from the splitting of DPs of the parent
Hermitian Floquet system.}
\label{fig_1Dvector}
\end{figure*}
Note in passing that, according to Eq.~(\ref{quasi_energy}), the
quasi-energy spectrum of our 2D model is determined by
\begin{eqnarray}
\cos \left[ \varepsilon \left( k_{x},k_{y}\right) \right] &=&\cos \left[
2t_{1}\left( \cos k_{x}+\cos k_{y}\right) +t_{2}\right]  \notag \\
&&\times \cos \left( 4t_{3}\sin k_{x}\sin k_{y}+i\gamma \right) ,
\end{eqnarray}%
where we set lattice constant, and driving period all equal to $1$. For
either $H_1$ or $H_2$ alone, one can see $H_{i}\left( \mathbf{k}\right)
=H_{i}\left( -\mathbf{k}\right) $ for all $\mathbf{k}$. Thus, $H_1$ or $H_2$
alone as a static system does not possess NHSE. That also indicates that in
the high frequency limit, the Floquet effective Hamiltonian becomes $(H_1+H_2)/2$ (with $T_1=T_2$)
and does not possess NHSE either. In general, by periodically quenching
between these two Hamiltonians, we may induce some loop structure in the
spectrum of the resulting Floquet effective Hamiltonian and hence FNHSE.
This will be elaborated in the next section using a reduced model.

{The Floquet EPs can now be determined by the following equations}
\begin{eqnarray}  \label{EP_2D2}
4t_{3}\sin k_{x}\sin k_{y} &=&n\pi ,  \label{EP_2D1} \\
2t_{1}\left( \cos k_{x}+\cos k_{y}\right) +t_{2} &=&m\pi \pm \arccos \left(
\frac{1}{\cosh \gamma }\right) ,  \notag \\
\end{eqnarray}%
where $m$, $n$ are integers of the same parity (opposite parity) if the EPs
locates at $\varepsilon \left( k_{x},k_{y}\right) =0$ $\left[ \varepsilon
\left( k_{x},k_{y}\right) =\pi \right] $. These specific solutions give many
EPs, as shown in Fig.~\ref{fig_split} (for simplicity, we set $T_{1}=T_{2}=1$
throughout this paper). Regarding how the presence of the non-Hermitian term
affects the parent Hermitian Floquet system, two important observations are
in order. First, the $i\gamma $ term can induce the splitting of a single DP
into two EPs. Each pair of the EPs denoted by either red star or blue circle
stemming from a black DP as determined by $2t_{1}\left( \cos k_{x}+\cos
k_{y}\right) +t_{2}=m\pi $ and $4t_{3}\sin k_{x}\sin k_{y}=n\pi $. This is
plotted in Fig.~\ref{fig_split}, where different colors are also used to
represent quasi-energy values at zero or $\pi$. In principle, each EP
inherits half of the topological winding number of the DP associated with
the parent Hermitian Floquet system. These EPs cannot be annihilated unless
two EPs with same quasi-energy collide with each other and therefore the
presence of EP is topological. Second, as compared with the Hermitian case,
the $i\gamma $ term can generate new intersections between the two closed
curves described by Eq.~(\ref{EP_2D2}) and Eq.~(\ref{EP_2D1}), thus creating
brand-new second-type EPs unrelated to the DPs of the parent Hermitian
system. This second type of EPs are denoted by diamonds in Fig.~\ref%
{fig_split}.  It is seen that the second-type of EPS always emerge in pairs.

To further characterize the vorticity of the EPs in the quasi-energy
spectrum, we next plot the planar vector field $\mathbf{D}_{\alpha }\left(
\mathbf{k}\right) $ in Fig.~\ref{fig_vector}. The vorticity of the EPs can
be quantified by the winding number $w_{\mathrm{I},j}$. To that end, we
first choose a closed loop in the momentum space, then we obtain a closed
loop in the vector field space. Similar to the Ref.~\cite{Lee}, $w_{\mathrm{I%
},j}$\ will be a half integer if the momentum-space loop encircles an EP of
the first type), and it must be zero if the loop does not encircle any EP of
the first type. It is also found that each pair of the second type of EPs, if both enclosed by the chosen momentum-space closed loop,
always yield $w_{\mathrm{I},j}=0$. Thus, the two members of each pair of the second-type of EPs always have opposite values of $w_{\mathrm{I},j}$.
All these results are independent of the quasi-energy of
the EPs $\left[ \varepsilon \left( \mathbf{k}\right) =0,\pi \right] $ and is
irrespective of the precise shape of the winding loop $C$.

To further illustrate our observations above, we now look into the
trajectory of $\left\langle \sigma _{x}\right\rangle _{1}$, $\left\langle
\sigma _{z}\right\rangle _{1}$ as $\mathbf{k} $ encircles the EPs found in
Fig.~\ref{fig_trajectory}. There it is seen the trajectory is half of a
closed curve, thus indicating that their winding number is one half. By
contrast, when $\mathbf{k} $ encircles both EPs with the same topological
charge, the resulting trajectory forms a closed curve and therefore the
corresponding winding number is $1$.

\section{Floquet Non-Hermitian skin effect and topological characterization}

\subsection{EPs in a 1D non-Hermitian Floquet system}

\label{1D}
\begin{figure*}[tbp]
\centering
\includegraphics[width=0.42\textwidth]{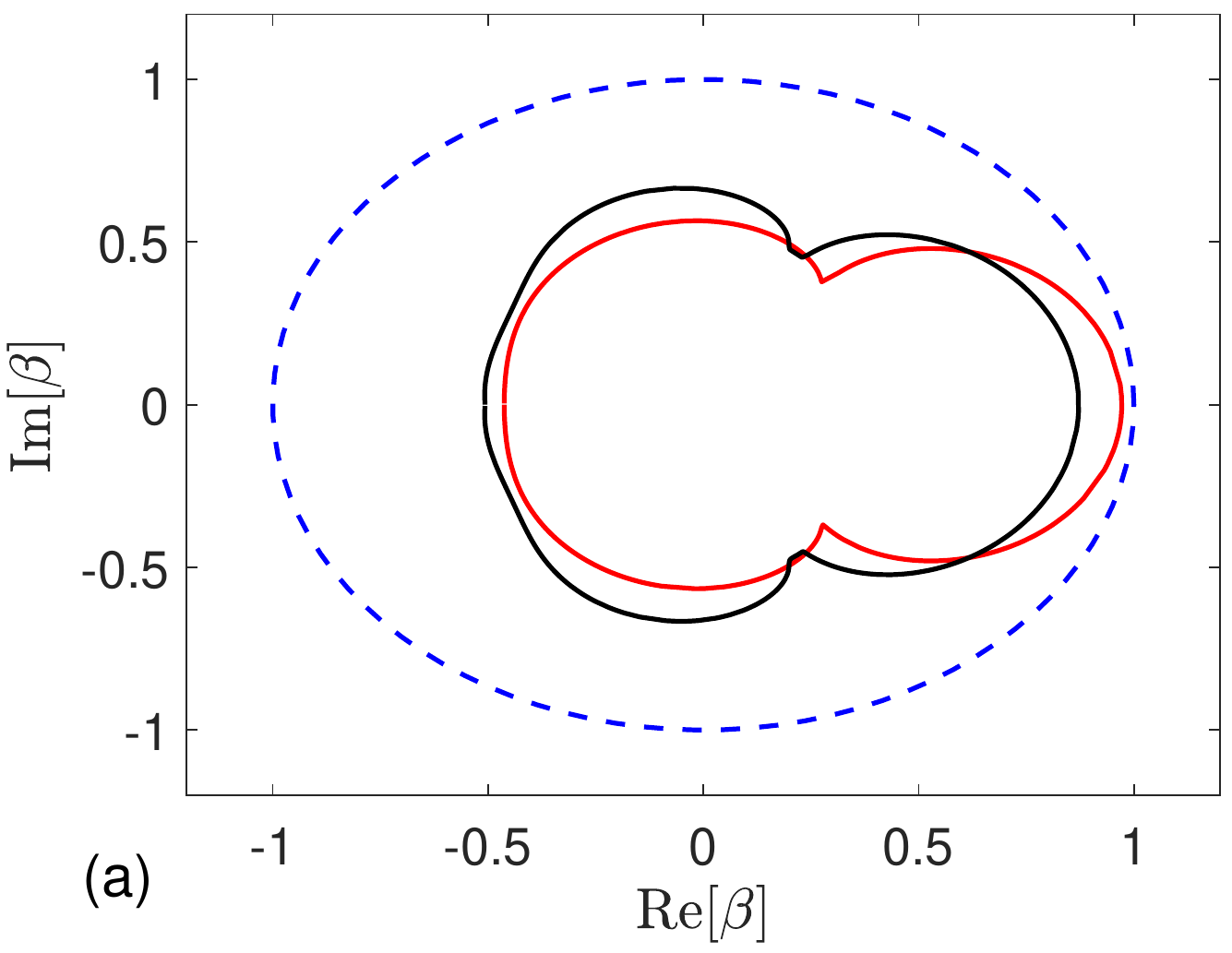} \includegraphics[width=0.42%
\textwidth]{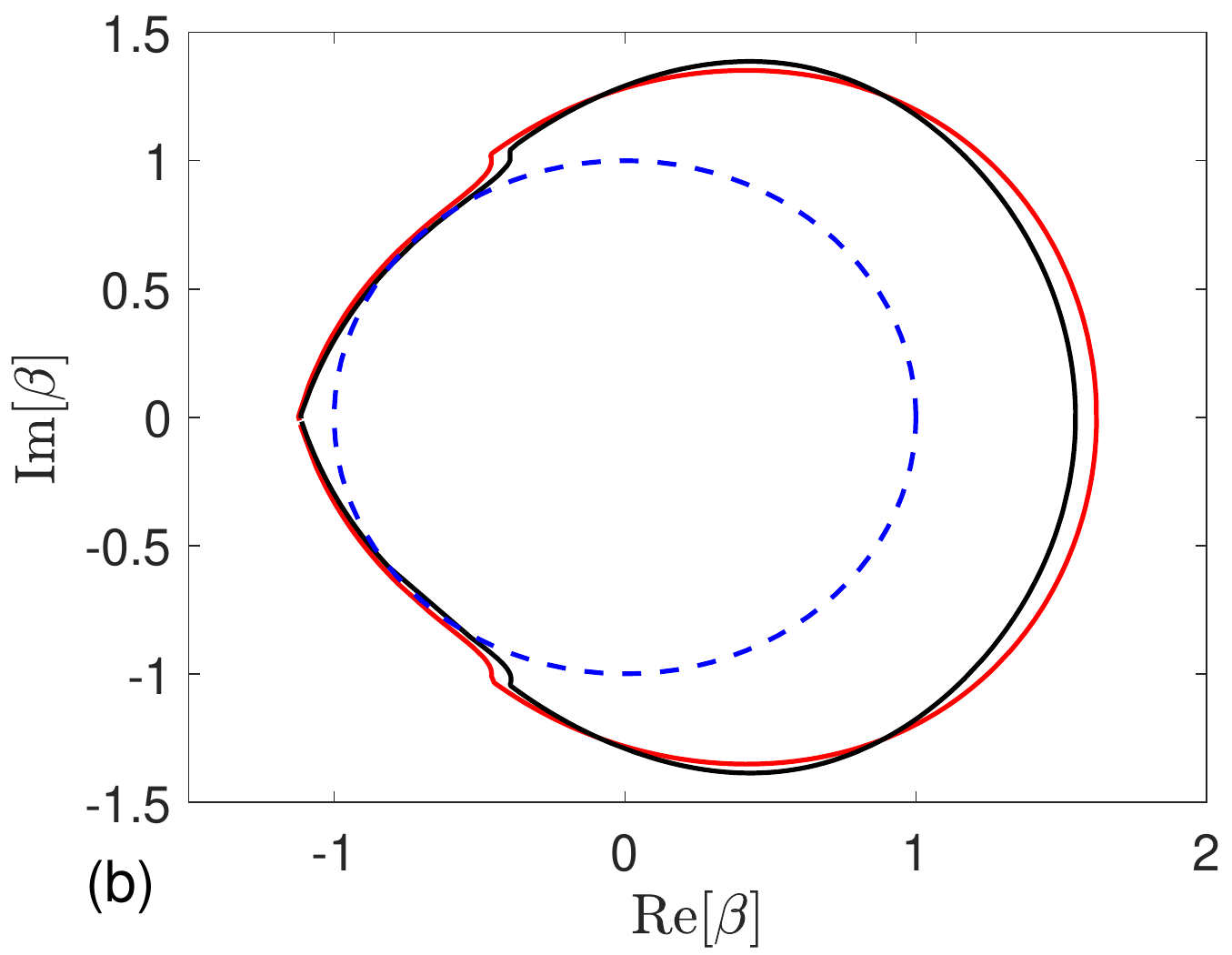}
\caption{GBZs, denoted by $C_{\protect\beta }$, obtained for two different
time-symmetric frames (with different colors). Details are specified in the
main text and in Appendix. (a) $w=0.5$, $v=0.8,$ and $\protect\gamma =0.5$.
(b) $w=2$, $v=0.6,$ and $\protect\gamma =0.5$. For comparison, we also plot $%
C_{k}$ with blue dashed lines in the absence of the non-Hermitian term. It
is seen that $|\protect\beta| $ can be larger or less than unity.}
\label{fig_gbz}
\end{figure*}
As become clear from explicit calculations presented in Appendix, it is
highly non-trivial to examine FNHSE in Floquet systems, even for 1D
situations. Recognizing this challenge, in this section we start with a 1D
non-Hermitian Floquet system periodically quenched between the following two
Hamiltonians,
\begin{eqnarray}
H_{1} &=&\sum_{k}B_{1,x}\left( k\right) \sigma _{x}\left\vert k\right\rangle
\left\langle k\right\vert , \\
H_{2} &=&\sum_{k}B_{2,z}\left( k\right) \sigma _{z}\left\vert k\right\rangle
\left\langle k\right\vert ,
\end{eqnarray}%
where
\begin{eqnarray}
B_{1,x}\left( k\right) &=&w+v\cos k \\
B_{2,z}\left( k\right) &=&v\sin k+i\gamma
\end{eqnarray}%
with $w$ and $v$ are the intracell and intercell couplings. $i\gamma $
denotes the staggered on-site imaginary potential. Although the driven
Hamiltonian in each period is simple, the Floquet system can still induce
many non-trivial physical properties. This 1D Floquet Hamiltonian can be
also deemed as the equivalent Hamiltonian describing the subspace of the 2D
Floquet system in Eqs.~(\ref{H1_2D})-(\ref{H2_2D}) associated with one
particular quasi-momentum along the $y$ direction. In this sense, examining
1D Floquet systems not only provides some insights into low-dimensional
FNHSE but also captures important aspects of the edge states of certain 2D
non-Hermitian Floquet systems. The locations of EPs of this 1D model are
found to be determined from
\begin{eqnarray}
v\sin k &=&n\pi , \\
w+v\cos k &=&m\pi \pm \arccos \left( \frac{1}{\cosh \gamma }\right) .
\end{eqnarray}%
Again, $m$, $n$ are integers of the same parity (opposite parity) if the gap
closes at $\varepsilon \left( k\right) =0$ $\left[ \varepsilon \left(
k\right) =\pi \right] $. In Fig.~\ref{fig_1Dvector}, we plot the obtained
EPs with different quasi-energies in the $w-k$ space. We choose the $w-k$
space to plot because, now given only one quasi-momentum variable, we are
forced to use a second system parameter to form a 2D parameter space so as
to define winding numbers etc. For example, the winding numbers $w_{\mathrm{I%
},j}$ now defined in the $w-k$ space also capture the topological features
of the EPs. As a specific example, we may examine $w_{\mathrm{I},j}$ on a
circle around the EPs, defined as $k=r_{1}\sin \theta +k_{c}$ and $%
w=r_{2}\cos \theta +w_{c}$ with $\theta $ varying from $0$ to $2\pi $, $%
r_{i} $ the radius of the ellipse and $\left( k_{c}\text{, }w_{c}\right) $
the coordinate of the EP in this parameter space. According to the Ref. \cite%
{Lilinhu}, the winding number can be therefore represented as
\begin{equation}
\sum_{i}w_{\mathrm{I},\alpha }^{i}=v_{L,\alpha }-v_{R,\alpha },
\label{winding_difference}
\end{equation}%
where
\begin{equation}
v_{L(R),\alpha }=\frac{1}{2\pi i}\int\nolimits_{k}\mathrm{d}k\text{ }\Xi
_{\alpha ,S}^{-1}\text{ }\frac{\mathrm{d}\Xi _{\alpha ,S}}{\mathrm{d}k},%
\text{ }\alpha =1,\text{ }2
\end{equation}%
denoting the winding numbers of different topological phases on two sides of
the EP. $w_{\mathrm{I},\alpha }^{i}$ represents the winding number of the
i-th EPs. In our 1D system shown in Fig.~\ref{fig_1Dvector}, only one EP is
available for a given value of $w$. Therefore we can remove the summation in
the Eq. (\ref{winding_difference}). It is clear that the change of winding
number $v_{\alpha }$ across the phase transition point equals to the
summation of winding number $w_{\mathrm{I},\alpha }^{i}$ of each EPs. From
this point of view, EP is also the topological phase transition point.
\begin{figure}[tbp]
\centering
\includegraphics[width=0.45\textwidth]{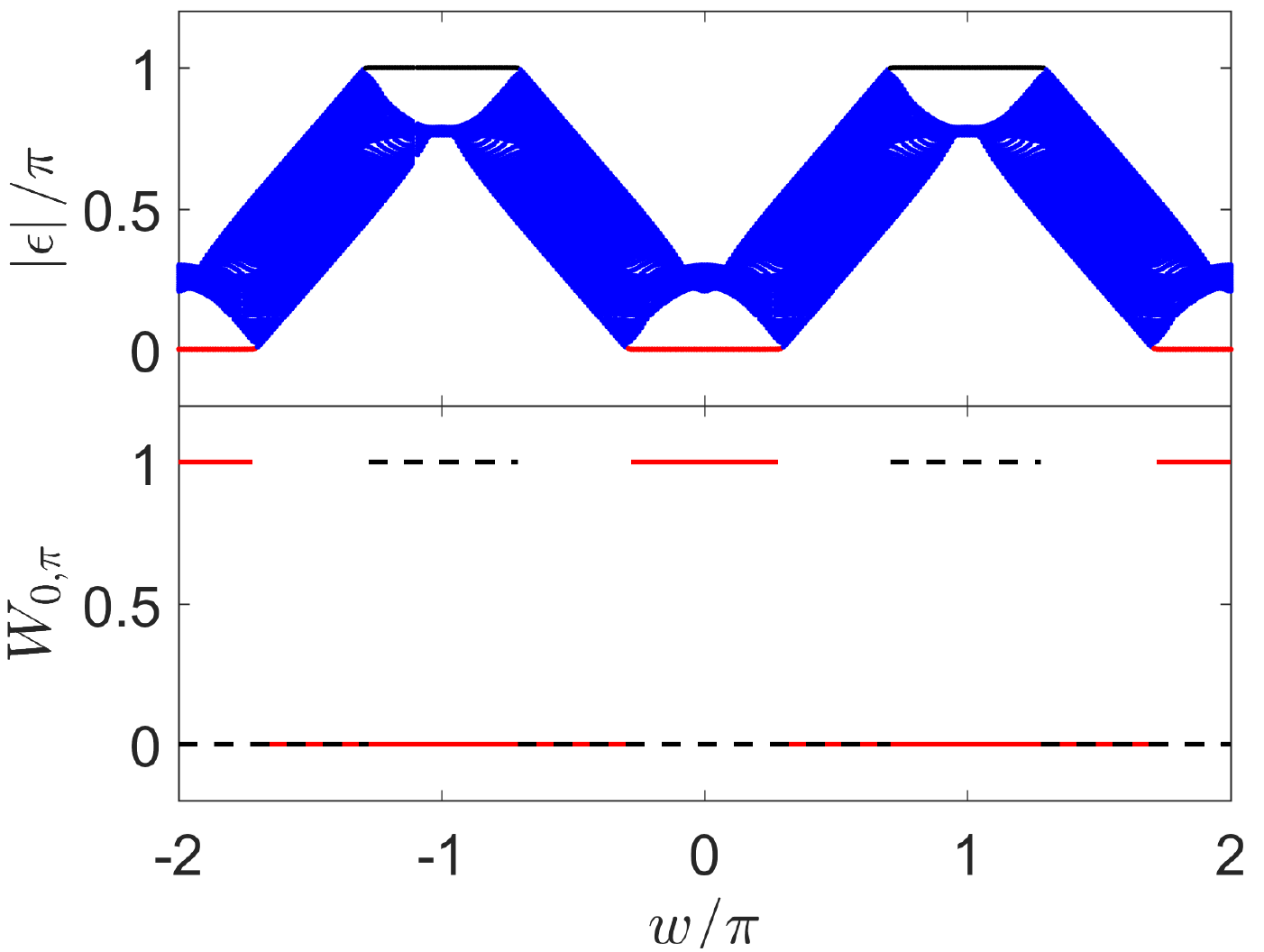}
\caption{The restored BBC for our 1D non-Hermitian Floquet system depicted
in the main text. (a) Upper panel: The OBC spectrum with $v=0.8$, and $%
\protect\gamma =0.5$. (b) Lower panel: The topological bulk invariants
calculated with the aid of GBZ. Red and blue points denote the winding
numbers $W_{0}$ and $W_{\protect\pi }$, respectively. The obtained winding
numbers can perfectly predict two types of Floquet edge modes.}
\label{fig_bbcwinding}
\end{figure}

\subsection{Skin effect and topological characterization}

\label{OBC}

For the present 1D non-Hermitian Floquet system, the Floquet operators in
two time-symmetric frames are given by%
\begin{equation}
U_{j}\left( k\right) =n_{0}\left( k\right) +i\mathbf{n}_{j}\left( k\right)
\cdot \mathbf{\sigma }\text{, }j=1,\text{ }2
\end{equation}%
where $j=1$, $2$, and the effective vector fields $\mathbf{n}_{j}\left(
k\right) $ are found to be
\begin{eqnarray}
n_{1,x}\left( k\right) &=&\sin \left[ B_{1,x}\left( k\right) \right] \cos %
\left[ B_{2,z}\left( k\right) \right] ,  \label{Fle1} \\
n_{1,z}\left( k\right) &=&\sin \left[ B_{2,z}\left( k\right) \right] ,
\label{Fle2} \\
n_{2,x}\left( k\right) &=&\sin \left[ B_{1,x}\left( k\right) \right] ,
\label{Fle3} \\
n_{2,z}\left( k\right) &=&\cos \left[ B_{1,x}\left( k\right) \right] \sin %
\left[ B_{2,z}\left( k\right) \right] .  \label{Fle4}
\end{eqnarray}%
Apparently, if $w\neq m\pi $, then the $w$ value drops out from expressions $%
\cos[B_{1,x)}(k)]=\cos[w+v\cos(k)]=\pm\cos[v\cos(k)]$ and $\sin[B_{1,x}(k)]=%
\sin[w+v\cos(k)]=\pm \sin[v\cos(k)]$ and the resulting Floquet operator
reduces to a previously studied model that is known to have no FNHSE \cite%
{ZhoulPRB}(this will be confirmed again below). In general situations, due
to FNHSE, the Floquet eigenstates do not necessarily extend over the bulk.
Instead, they can be localized at either end of the lattice. The bulk
topological invariants [Eq. (\ref{zero_and_pi})] in terms of the Bloch wave
vector can no longer predict exactly the number of edge modes under OBC. One
main purpose of a careful topological characterization for our 1D model is
to inspect the possibility of restoring BBC in the presence of FNHSE. This
presents a challenge given the dramatic difference between the PBC spectrum
and the OBC spectrum caused by FNHSE.

To attack this issue below we shall
follow a generalized Bloch band theory by obtaining first the so-called the
generalized Brillouin Zone (GBZ) \cite{Yao1,Yokomizo}. In short, we aim to
recover the BBC by accounting for the non-Bloch-wave character of bulk
states through GBZ and then recalculate topological invariants based on GBZ.
\begin{figure}[tbp]
\centering
\includegraphics[width=0.42\textwidth]{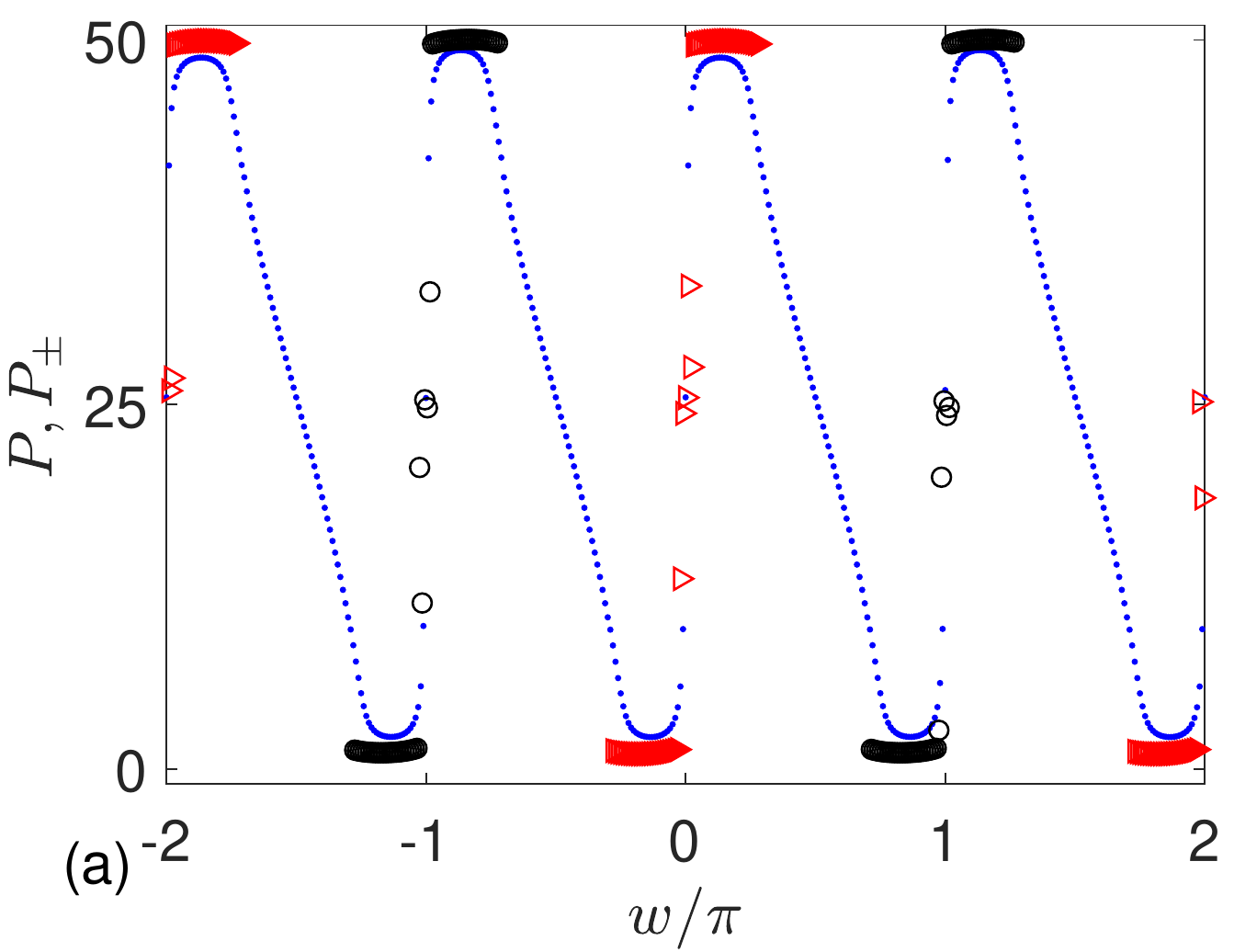} %
\includegraphics[width=0.42\textwidth]{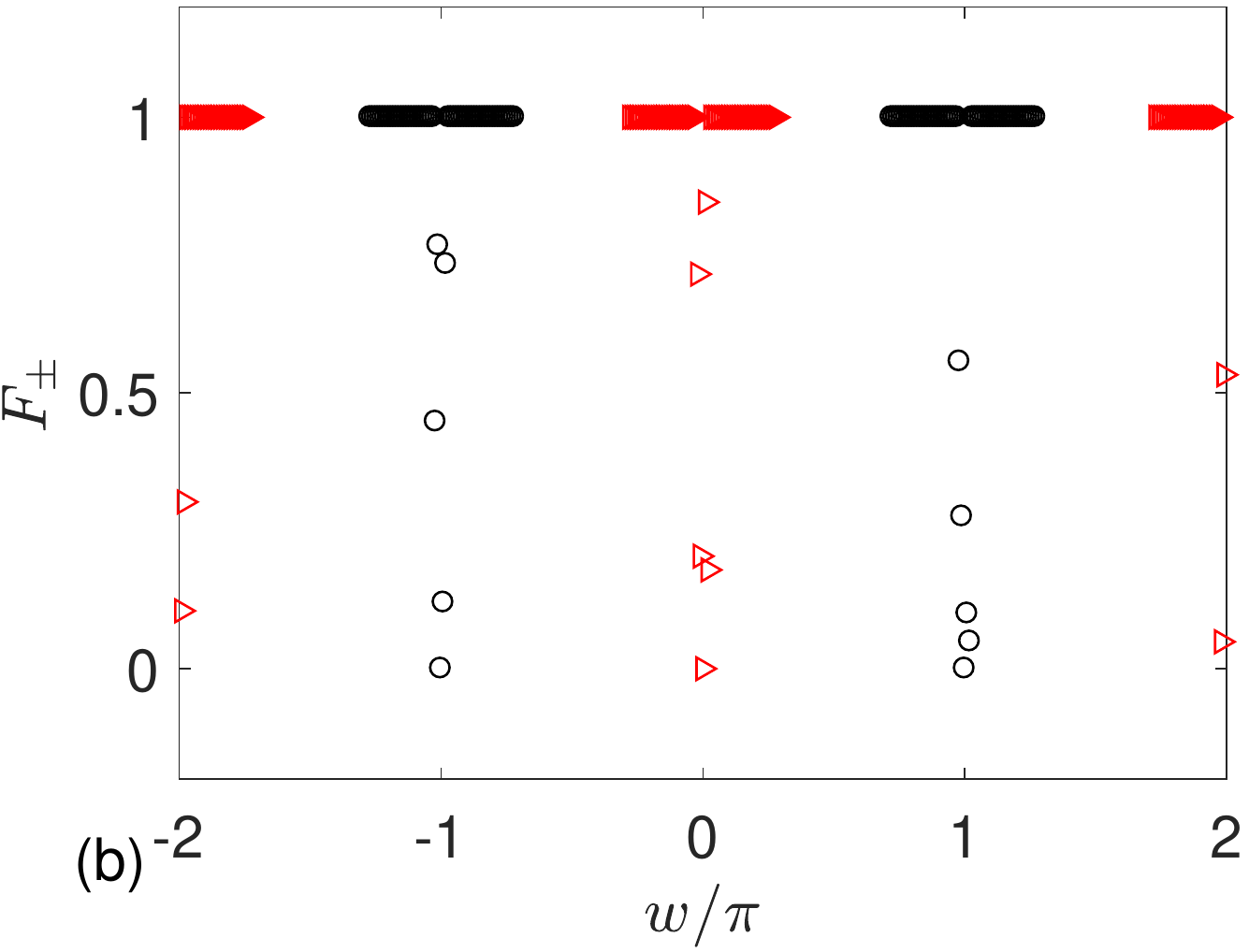}
\caption{Plots of the physical quantites of average position $P$ ($P_{\pm }$%
), and fidelity $F_{\pm }$ as functions of $w$. The numerical simulation is
performed for $v=0.8$, and $\protect\gamma =0.5$. (a) Blue dots, red
triangles, and black circles denote $P$ , $P_{+}$ and $P_{-}$ respectively.
(b) Red triangles and black circles represent $F_{+}$ and $F_{-}$
respectively. It can be seen that FNHSE can push all the Floquet eigenstates
to one side of lattice accompanied by the coalescence of the Floquet edge
modes, which can be shown in panel (b). Note that the fidelity $F_{\pm }$
quickly drops from $1$ to $0$ as the system parameter $w$ approaches $m
\protect\pi$. At these special points, the Floquet system is free of FNHSE
and therefore the edge modes localize individually at both ends of the
lattice.}
\label{fig_statefidelity}
\end{figure}
In the generalized Bloch band theory \cite{Yao1,Yokomizo}, one key step is
to replace $\exp \left( \pm ik\right) $ with $\beta ^{\pm 1}$, which can be
determined by the characteristic polynomial Det$\left[ H_{\mathrm{eff}%
}^{i}\left( \beta \right) -\varepsilon I\right] =0$ and other conditions.
For non-Hermitian systems in general, $\left\vert \beta \right\vert $ is not
unity and the corresponding bulk states are localized at one boundary \cite%
{Yao1,Chinghua,Yokomizo}. Specifically, if we number the solutions to Det$%
\left[ H_{\mathrm{eff}}^{i}\left( \beta \right) -\varepsilon I\right] =0$ as
$\beta _{i}\left( i=1,...,2M\right) $ so as to satisfy $\left\vert \beta
_{1}\right\vert \leq \left\vert \beta _{2}\right\vert \leq \left\vert \beta
_{3}\right\vert \leq \left\vert \beta _{2M}\right\vert $, where $2M$ stands
for the degree of algebraic equation for $\beta $ \cite{Yokomizo}. The
so-called GBZ, denoted $C_{\beta }$, is obtained by all solutions satisfying
the curious condition $\left\vert \beta _{M}\right\vert =\left\vert \beta
_{M+1}\right\vert $ (see Appendix for more details).
\begin{figure}[tbp]
\centering
\includegraphics[width=0.45\textwidth]{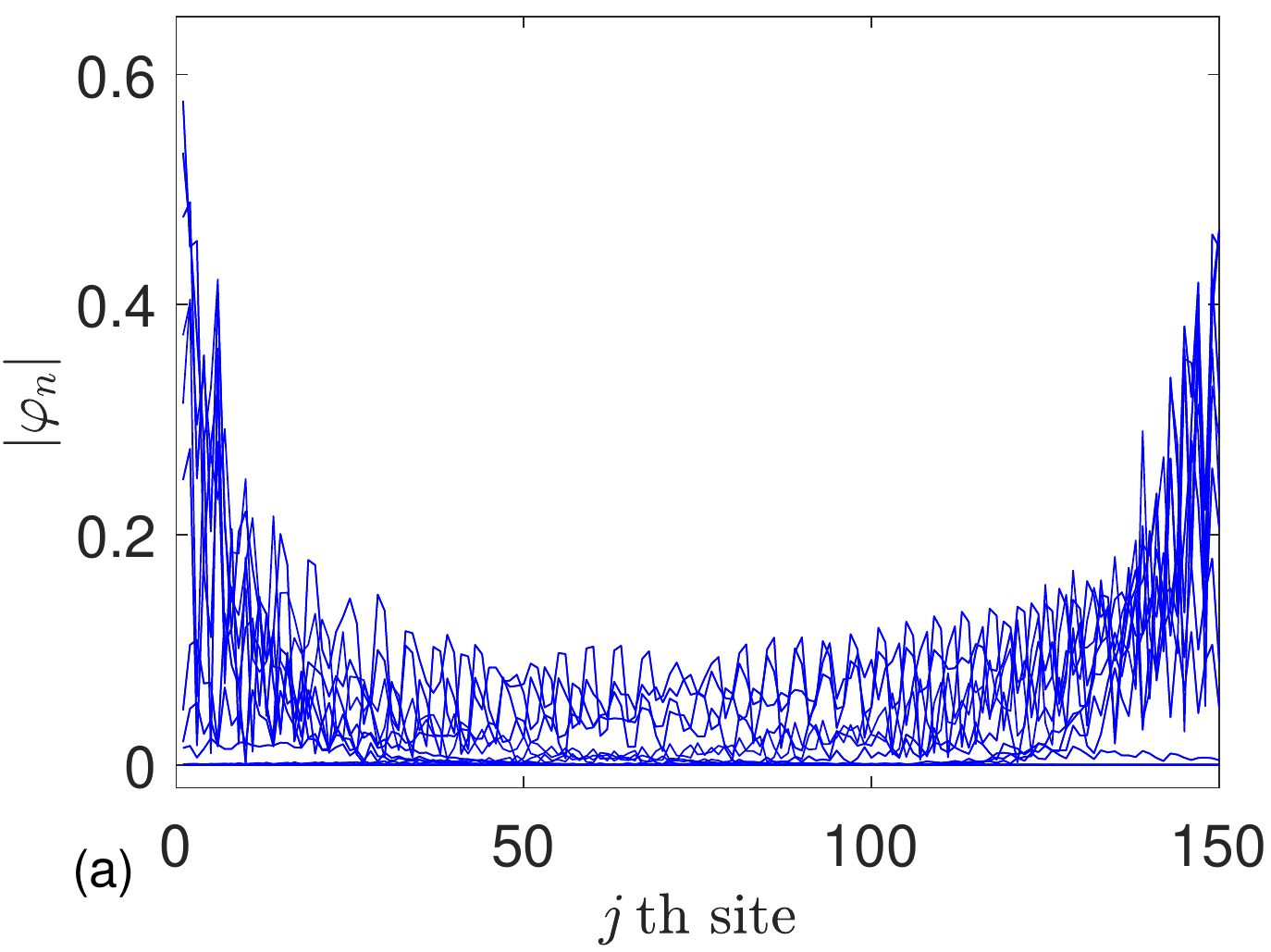} \vskip+5.2mm %
\includegraphics[width=0.45\textwidth]{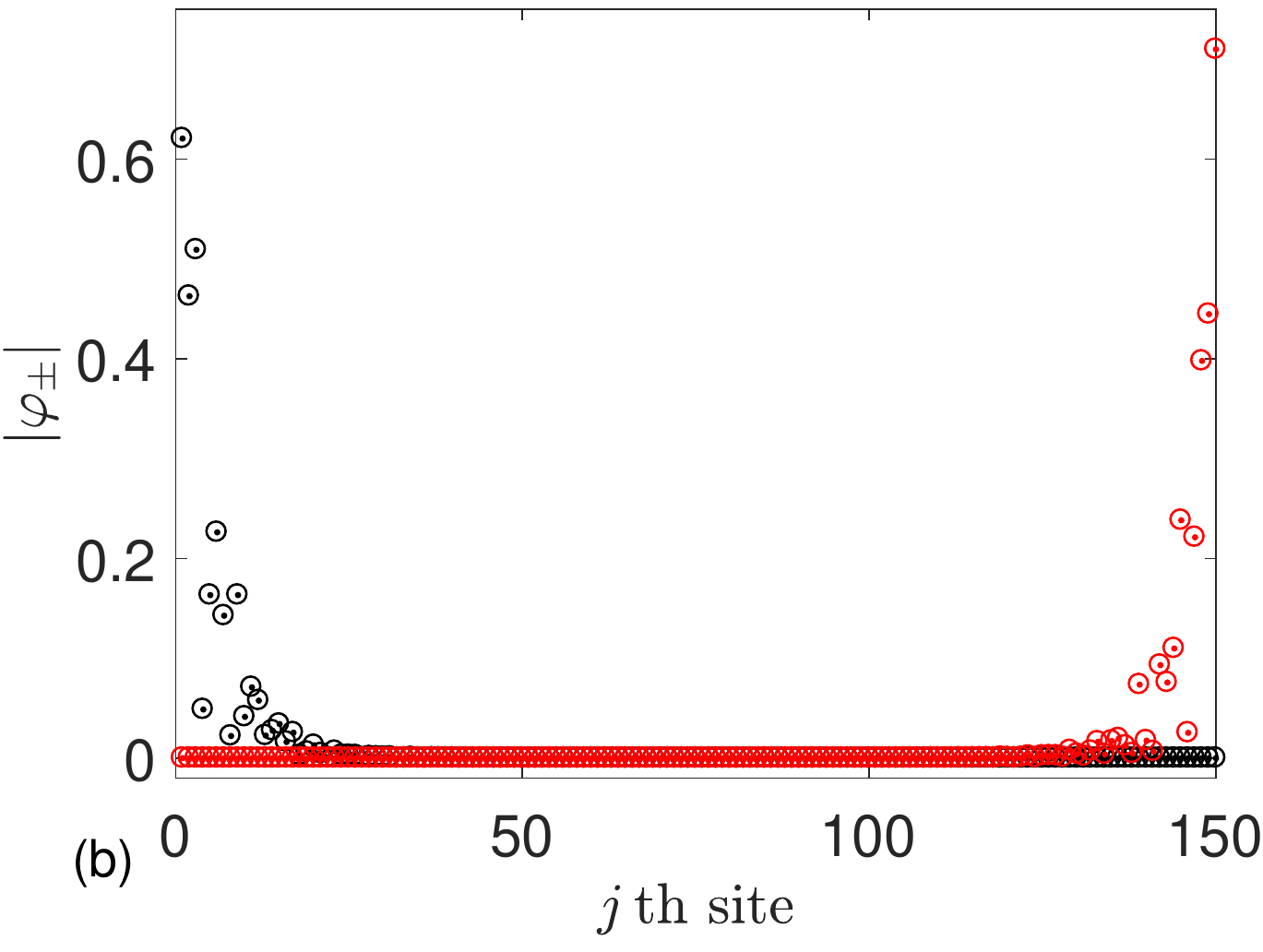}
\caption{{Profile of the eigenstates $\left\vert \protect\varphi %
_{n}\right\rangle $ $w=-4.9$, $v=1.8$, and $\protect\gamma =0.5$, for a lattice of 150 sites.   (a) Bulk
states denoted by blue lines. (b) Floquet edge states. Red (black) dots and
circles denote two $0$ edge ($\protect\pi $) modes, respectively. It can be
readily seen that the two edges modes coalesce individually and localize at
two different ends of the lattice.}}
\label{fig_quasistate}
\end{figure}

{One can now appreciate that consideration of GBZ in Floquet systems can be
highly technical. In static systems (or a Floquet system in the
high-frequency limit), the effective Hamiltonian $H_{\mathrm{eff}}^{j}$ only
contains some short-range couplings (e.g., nearest-neighbor hopping). Then
the equation Det$\left[ H_{\mathrm{eff}}^{i}\left( \beta \right)
-\varepsilon I\right] =0$ involves only a finite-order polynomial of $\beta $%
. However, even in our simple 1D Floquet system here, the periodic quenching
effectively induces long-range hopping across the lattice. This fact can be
also seen from the momentum-space function $\sin \left[ \cos \left( .\right) %
\right] $ contained in Eqs.~(\ref{Fle1})-(\ref{Fle4}), which can be directly
interpreted as a consequence of effectively long-range hopping across the
lattice. This hence leads to a highly complicated characteristic equation Det%
$\left[ H_{\mathrm{eff}}^{i}\left( \beta \right) -\varepsilon I\right] =0$.
Indeed, for our 1D model here, Det$\left[ H_{\mathrm{eff}}^{i}\left( \beta
\right) -\varepsilon I\right] =0$ involves functions such as $\sin \left[
w+v \cos \left( \beta/2+1/2\beta\right) \right]$, which is a polynomial of $%
\beta$ to infinite order. }

To provide part of the solutions, we now assume that the strength of $v$ is
less than $1$ such that the involved polynomials about $\beta$ can be
effectively truncated up to $\beta^2$ and $\beta^-2$ (for more details, see Appendix).
This procedure also restricts ourselves to a parameter regime without the
coexistence of many pairs of Floquet edge states. In Fig.~\ref{fig_gbz}, we
plot the GBZ according to our calculations detailed in Appendix. It can be
shown that $\left\vert \beta \right\vert $ can be larger or smaller than $1$%
. Furthermore, $C_{\beta }$ can have cusps at which more than two solutions
of $\beta $ share the same absolute value \cite{Yokomizo}.

To proceed, we replace the $C_{\mathbf{k}}$ with $C_{\mathbf{\beta }}$ in
Eq.~(\ref{w2}). We then calculate the generalized winding numbers $W_{0}$
and $W_{\pi }$ in terms of the reduced vector $\mathbf{n}_{i}\left(
\beta\right) $. Presented in Fig. \ref{fig_bbcwinding} is the OBC spectrum
of our non-Hermitian system and the corresponding generalized winding
numbers. Excellent agreement is obtained and the BBC is perfectly
recovered for small values of $v$. Note also that even when we only
approximate the characteristic equation Det$\left[ H_{\mathrm{eff}%
}^{i}\left( \beta \right) -\varepsilon I\right] =0$ up to the second order,
BBC for our 1D Floquet system can already be well recovered. One may perform similar calculations for based on a third-order expansion in terms of $v$, but that is already more accurate than necessary in terms of
recovering the BBC here in the presence of FNHSE,

\subsection{Anomalous $0$ and $\protect\pi $ modes under OBC}

\label{edge_modes}

In this subsection, we investigate the Floquet eigenstates under OBC. It can
be envisioned that the Floquet eigenstates will be pushed to one lattice
boundary due to FNHSE. However, because of the absence of FNHSE for $w=m \pi$%
, we anticipate some interesting details as we tune $w$ across these special
points. In addition, for systems with larger and larger $v$, the emergence
of longer-range hopping can make the situation richer, restoration of BBC
will not be practical and this makes our investigation of the Floquet edge
states more necessary.

Let us first focus on the case of with small $v$, namely $v=0.8$ as an
example. At $w=m\pi $, the system is free from FNHSE. What we are curious
about is how the Floquet eigenstates change when the system crosses such
FNHSE-free points, and whether this transition is different for bulk and
edge modes. To that end we define several quantities. We first define $P$
and $P_{\pm }$ as follows:
\begin{equation}
P=\frac{1}{N}\sum_{n}\frac{\left\langle \varphi _{n}\right\vert \widehat{x}%
\left\vert \varphi _{n}\right\rangle }{\left\langle \varphi _{n}\right.
\left\vert \varphi _{n}\right\rangle },\text{ }P_{\pm }=\frac{1}{N_{\mathrm{%
edge}}}\sum_{j}\frac{\left\langle \varphi _{\pm ,j}\right\vert \widehat{x}%
\left\vert \varphi _{\pm ,j}\right\rangle }{\left\langle \varphi _{\pm
,j}\right. \left\vert \varphi _{\pm ,j}\right\rangle },
\end{equation}%
where $\left\vert \varphi _{n}\right\rangle $ represents the $n$th bulk
eigenstates, and $\left\vert \varphi_{\pm, j}\right\rangle $ represents the $%
j$th Floquet edge mode at quasi-energy $0$ (for $+$) and $\pi$ (for $-$),
respectively. $N$ ($N_{\mathrm{edge}}$) is the number of of the bulk (edge)
states, and $\widehat{x}$ is the position operator. $P$ therefore measures
the average position of the bulk states, and $P_{\pm }$ measures the same
for Floquet edge states at quasi-energy $0$ and $\pi$.
\begin{figure*}[tbp]
\centering
\includegraphics[width=0.42\textwidth]{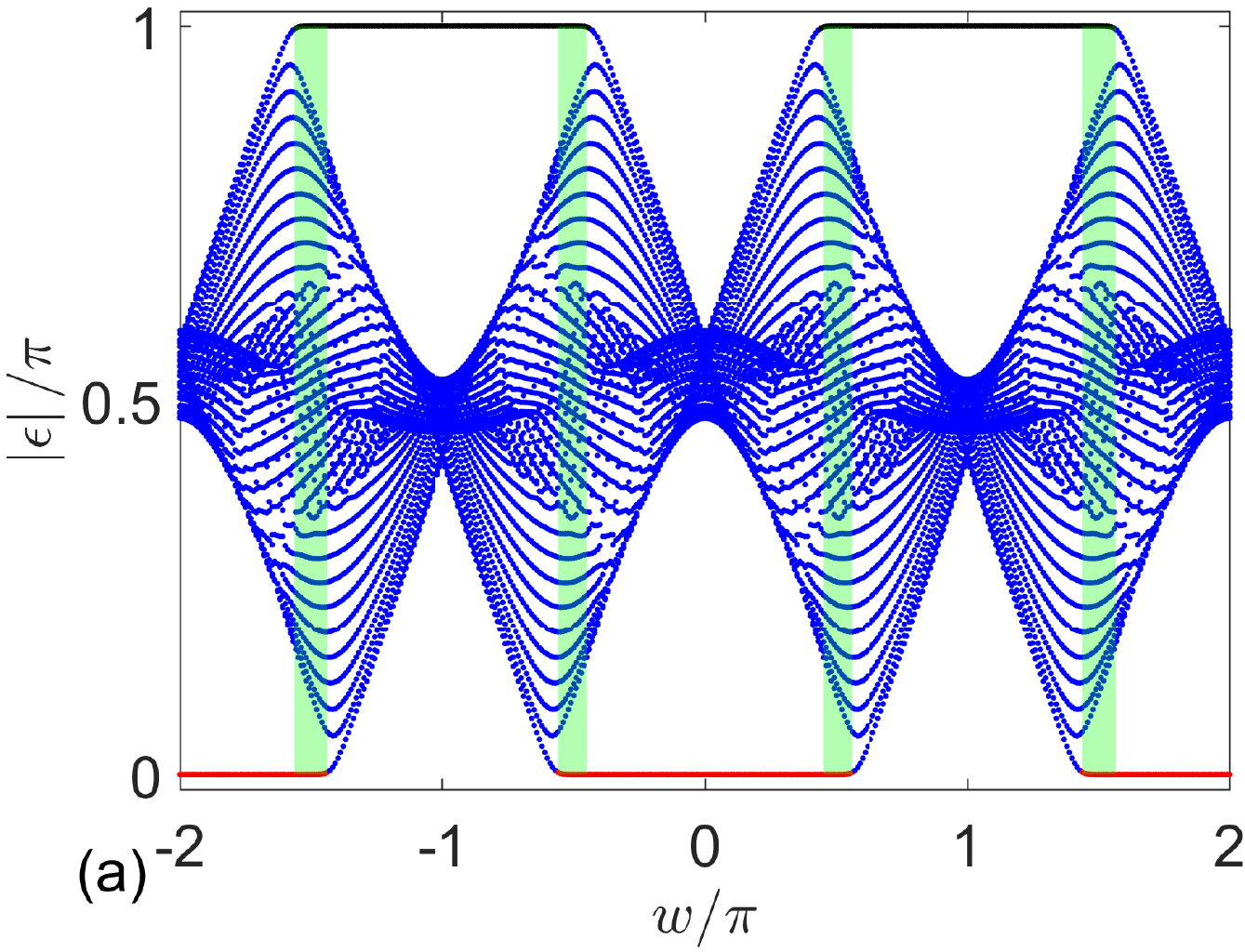} %
\includegraphics[width=0.43\textwidth]{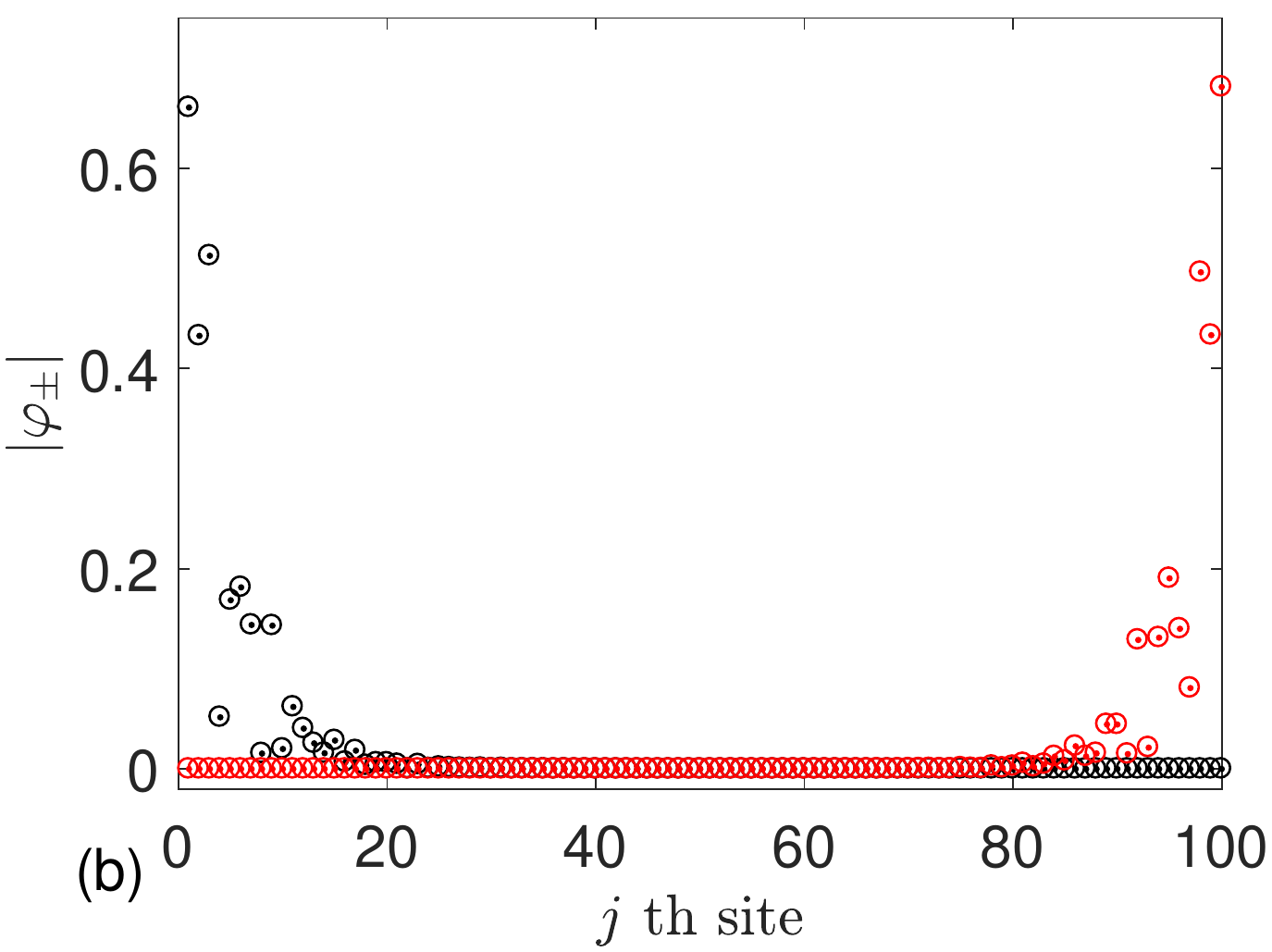}
\caption{Coexistence of two types of Floquet edge states. (a) The OBC
spectrum of our non-Hermitian Floquet system $v=1.8 $, and $\protect\gamma %
=0.5$, for a lattice of 100 sites.  The four green shaded regions indicate the coexistence of the $0$ and
$\protect\pi $ modes. (b) Typical profile of the two types of Floquet edge
modes. Red (black) dots and circles denote the two edge $0$ ($\protect\pi $)
modes, respectively. The Floquet edge modes localize at two different boundaries
of the lattice.}
\label{fig_coexistence}
\end{figure*}
%
If all the bulk eigenstates are extended states or all the edge modes come
in pair and are localized on both sides of the lattice, then the values of $%
P $ and $P_{\pm }$ will be half of the lattice length. The other quantity
describes the fidelity between two edge modes with the same quasi-energy:
\begin{equation}
F_{\pm }\left( j,j^{\prime }\right) =\left|\left\langle \varphi _{\pm
,j}\right. \left\vert \varphi _{\pm ,j^{\prime }}\right\rangle\right| .
\end{equation}%
If $F_{\pm }\left( j,j^{\prime }\right) =1$, then the corresponding two edge
modes coalesce. Below we call such coalescence edge modes as anomalous
Floquet edge modes.

Figure~\ref{fig_statefidelity} shows how FNHSE impacts on the Floquet
eigenstates in terms of the quantities defined above. There it is seen that
FNHSE pushes all Floquet eigenstates to one boundary. In particular, except
for cases with $w$ very close to special points of $w=m\pi$, all Floquet
edge modes coalesce and are localized on only one side of the lattice. The
values of $P$ and $P_{\pm }$ undergo dramatic changes as $w$ crosses the
skin-effect-free points. As such, the average location of both the bulk
states and coalescent edge states switches from one side to the other side
of the lattice. The preferred direction of FNHSE here can thus be tuned via
the system parameter $w$.

Next we switch to cases with larger $v$, i.e., $v=1.8$. This value is
already beyond the regime where our theoretical treatment can restore BBC.
Results are presented in Fig.~\ref{fig_quasistate}, in terms of the
populations on each lattice sites for individual eigenstates under OBC. It
can now be seen that FNHSE pushes most of the Floquet eigenstates to both
lattice boundaries instead of one boundary. Thus, FNHSE for larger values of
$v$ is richer as its preferred direction changes from one bulk eigenstate to
another. Furthermore, the two types of edge modes ($0$ and $\pi $) can now
coexist under OBC spectrum, (see green rectangles of Fig. \ref%
{fig_coexistence}). Remarkably, we find that in the present parameter regime
each of the coexisting edge modes coalesces individually and localizes at
two different boundaries of the lattice. If we sweep the parameter $w$ across $%
m\pi $, then the Floquet zero modes and $\pi$ modes will also exchange their
preferred boundaries. To our knowledge, such anomalous localization
phenomena are not known previously and it would be stimulating to develop a
theory to account for these findings. The separation of Floquet zero modes
and $\pi$ modes by FNHSE may be also beneficial when considering dynamics
control, information encoding and even braiding between Floquet zero modes
and $\pi$ modes \cite{Bomantara1,Bomantara2}.

\section{Conclusions}

\label{conclusions} In this work, we have investigated some general aspects
of periodically driven non-Hermitian lattice systems in 1D and 2D. For
Floquet band structure under PBC, we focus on EPs and their topological
characterization. For Floquet spectrum under OBC, we place our emphasis on
FNHSE and the BBC. Though our explicit results are based on simple
periodically quenched two-band models, we can still make a number of general
conclusions, as listed below.

First, Non-Hermiticity in Floquet systems can induce many EPs. One type of
EPs is inherited from the DPs of their parent Hermitian Floquet system with
half-integer topological winding numbers. The second type of EPs, newly
created by non-Hermiticity, always appear in pairs and have opposite
half-integer winding numbers. The number and the locations of such EPs can
be extensively controlled by tuning some system parameters.

Second, non-Hermitian Floquet systems in general have FNHSE, albeit with
some special FNHSE-free points. This explains why in a previous study \cite%
{ZhoulPRB}, FNHSE was not found. The preferred direction of FNHSE can be
controlled if we tune relevant system parameters across such FNHSE-free
points. This observation may lead to new opportunities for quantum control
with the aid of non-Hermiticity.

Third, in the presence of FNHSE, BBC breaks down as expected. Recovering BBC
can in principle be done in Floquet systems, in the same fashion as in
static systems using, for example, the GBZ framework. However, even for our
model systems whose bulk spectrum under PBC can be worked out analytically,
restoration of BBC is only feasible in practice for certain parameter
regimes where low-order truncation of the characteristic polynomial can be
done with negligible error. As such, in general, predicting the Floquet edge
states by use of the bulk properties may present a challenge.

Fourth, to motivate further studies, a few interesting aspects of
non-Hermitian Floquet edge states deserve to be highlighted. In particular,
Floquet zero and $\pi $ modes are found to be coalescent modes in general.
Their existence can be well predicted from topological winding numbers of
the bulk states if BBC can be restored. These two types of coalescent edge
modes can also coexist (in our model system, it is already highly
challenging to predict this coexistence using restored BBC). When they do
coexist, they can be localized at different boundaries of the system. It is
worth pointing out that such characteristics are very different from those
in the work \cite{ZhoulPRB} in which BBC holds and many non-coalescent edge
modes can exist due to the absence of FNHSE.

\acknowledgments We are grateful to Ching-Hua Lee, Linhu Li, and Longwen
Zhou for stimulating discussions. J.G. is supported by the Singapore NRF
grant No.~NRF-NRFI2017-04 (WBS No.~R-144-000-378-281). X.Z. was supported by
National Natural Science Foundation of China (Grant No. 11975166) before he
joined NUS.

\appendix
\label{appendix}

\section{Generalized Brillouin Zone in two symmetric time frames}

In this Appendix, we shall explain how we obtain the GBZ in two time-symmetric
frames mentioned in the main text. Let us start from the Floquet operator in the first time
frame, which can be written as

\begin{equation}
U_{1}\left( k\right) =n_{0}+i\mathbf{n}_{1}\left( k\right) \cdot \mathbf{%
\sigma }\text{, }
\end{equation}%
where $n_{1,x}\left( k\right) =\sin \left( w+v\cos k\right) \cos \left(
v\sin k+i\gamma \right) $, and $n_{1,z}\left( k\right) =\sin \left( v\sin
k+i\gamma \right) $. The first step is to replace $e^{ik}$ with $\beta $ so
that the effective magnetic field can be expressed as
\begin{eqnarray}
n_{1,x}\left( \beta \right) &=&\sin \left[ w+v\left( \beta +\beta
^{-1}\right) /2\right]  \notag \\
&&\times \cos \left[ -iv\left( \beta -\beta ^{-1}\right) /2+i\gamma \right] ,
\\
n_{1,z}\left( \beta \right) &=&\sin \left[ -iv\left( \beta -\beta
^{-1}\right) /2+i\gamma \right] ,
\end{eqnarray}%
The corresponding characteristic equation of $\mathbf{n}_{1}\left( k\right)
\cdot \mathbf{\sigma }$ can be written as
\begin{equation}
\varepsilon ^{2}\left( \beta \right) =n_{1,x}^{2}\left( \beta \right)
+n_{1,z}^{2}\left( \beta \right)
\end{equation}%
We should choose the values of $\beta $ such that the energy levels become
dense and asymptotically form continuum bands when the system size increases.
However, the challenge here is that the polynomial with respect to $\beta $ is of infinite
order. It is too cumbersome to arrive at the general solution to Det$\left[ H_{\mathrm{eff}%
}^{i}\left( \beta \right) -\varepsilon I\right] =0$.   To obtain an analytical result, we need to perform
the Taylor expansion of $\varepsilon \left( \beta \right) $ up to finite
order. This can be done by considering a small $v$ coupling. Then, we
can approximate the characteristic equation up to the second order (as an example here.  One may also try the same expansion up to the third order) with respect
to $v$. Under this treatment, the approximate expression of $\varepsilon \left(
\beta \right) $ can be given as
\begin{equation*}
\varepsilon ^{2}\left( \beta \right) =\sum_{j=-2}^{2}\beta ^{j}X_{j},
\end{equation*}%
where
\begin{eqnarray}
X_{2} &=&\frac{v^{2}}{16}\left[ 1+2\cos \left( 2w\right) -2\cosh ^{2}\left(
\gamma \right) \right.  \notag \\
&&\left. -\cosh \left( 2\gamma \right) -4\sin \left( 2w\right) \sinh \left(
2\gamma \right) \right] , \\
X_{1} &=&v\cos \left( 2w\right) \cosh \left( \gamma \right) \left[ \cosh
\left( \gamma \right) \sin \left( w\right) \right.  \notag \\
&&\left. +\cos \left( w\right) \sinh \left( \gamma \right) \right] , \\
X_{0} &=&\left[ 2+v^{2}+\left( 3v^{2}-2\right) \cos \left( 2w\right) \right]
\cosh ^{2}\left( \gamma \right)  \notag \\
&&+\left[ -4+v^{2}+v^{2}\cos \left( 2w\right) \right] \sinh ^{2}\left(
\gamma \right) , \\
X_{-1} &=&v\cos \left( 2w\right) \cosh \left( \gamma \right) \left[ \cosh
\left( \gamma \right) \sin \left( w\right) \right.  \notag \\
&&\left. -\cos \left( w\right) \sinh \left( \gamma \right) \right] , \\
X_{-2} &=&\frac{v^{2}}{16}\left[ 1+2\cos \left( 2w\right) -2\cosh ^{2}\left(
\gamma \right) \right.  \notag \\
&&\left. -\cosh \left( 2\gamma \right) +4\sin \left( 2w\right) \sinh \left(
2\gamma \right) \right] .
\end{eqnarray}%
This is a quartic equation equation with respect to $\beta$ and therefore $%
M $ is $2$. Note that the degree of algebraic equation for $\beta $ depends
on the order $n$ of our Taylor expansion here, i.e., $M=n$. Moreover, the trajectory
of $\beta $ satisfying the continuum band can be determined by $\left\vert
\beta _{2}\right\vert =\left\vert \beta _{3}\right\vert $. Suppose the two
solutions $\beta $ and $\beta ^{\prime }$ have the same absolute values,
then we have $\beta ^{\prime }=\beta e^{i\theta }$ where $\theta $ is a real
number. Taking the difference between two characteristic equations $%
\varepsilon ^{2}\left( \beta \right) =F\left( \beta \right) $ and $%
\varepsilon ^{2}\left( \beta ^{\prime }\right) =F\left( \beta ^{\prime
}\right) $, we have
\begin{equation}
0=\sum_{j=-2}^{2}\beta ^{j}X_{j}\left( 1-e^{ij\theta }\right) .
\end{equation}

This equation allows us to obtain $\beta $ for a given value of $\theta
\in \left[ 0\text{, }2\pi \right) $. Then we obtain a set of values of $\beta $
that satisfies $\left\vert \beta \right\vert =\left\vert \beta ^{\prime
}\right\vert $. Picking up $\beta $ from the constraint $\left\vert \beta
_{2}\right\vert =\left\vert \beta _{3}\right\vert $, we finally arrive at the GBZ. One
can follow the same method to obtain the GBZ associated with  the Floquet operator for the
second time frame.

\end{document}